\documentclass[draft,jgrga]{agu2001}
\usepackage{graphicx}
\authorrunninghead{mori ET AL.}

\titlerunninghead{two-dimensional burridge-knopoff model}

\authoraddr{T. Mori,
Department of Earth and Space Science, Faculty of Science,
Osaka University, Toyonaka 560-0043,
Japan. (mori@spin.ess.sci.osaka-u.ac.jp)}

\def\gsim{\buildrel {\textstyle >}\over {_\sim}}
\def\lsim{\buildrel {\textstyle <}\over {_\sim}}

\begin{document}

%
%
%
%
%

%
%

\title{Simulation study of the two-dimensional Burridge-Knopoff model of earthquakes}
%

%
%


\author{Takahiro Mori and Hikaru Kawamura}
\affil{Department of Earth and Space Science, Faculty of Science,
Osaka University, Toyonaka 560-0043,
Japan}

\begin{abstract}
Spatiotemporal correlations of the two-dimensional spring-block  (Burridge-Knopoff) model of earthquakes are  extensively studied by means of numerical computer simulations. The model is found to exhibit either a ``subcritical'', ``supercritical'' or ``near-critical'' behavior, depending on the values of the model parameters. Transition between the ``subcritical'' and ``supercritical'' regimes is either continuous or discontinuous. Seismic events in the ``subcritical'' regime and those in the ``supercritical'' regime  at larger magnitudes exhibit universal scaling properties. In the ``supercritical'' regime, eminent spatiotemporal correlations, {\it e.g.\/}, a remarkable growth of seismic activity preceding the mainshock, arise in earthquake occurrence, whereas such spatiotemporal correlations are significantly suppressed in the ``subcritical'' regime.  Seismic activity is generically suppressed just before the mainshock in a close vicinity of the epicenter of the upcoming event while it remains to be active in the surroundings (the Mogi doughnut). It is also observed that, before and after the mainshock, an apparent $B$-value of the magnitude distribution decreases or increases in the ``supercritical'' or ``subcritical'' regimes, respectively. Such distinct precursory phenomena may open a way to the prediction of the upcoming large event. 

\end{abstract}

%
%

%

\begin{article}

%
%

\section{Introduction}

 An earthquake is a stick-slip dynamical instability of a pre-existing
 fault driven by the motion of a tectonic plate [\textit{Scholz}, 2002;\textit{Scholz}, 1998]. 
While an earthquake is a complex phenomenon, certain empirical laws
such as the Gutenberg-Richter (GR) law and the Omori law concerning
its statistical properties are known to hold. 
These are both power-laws without any characteristic energy or time scale. Such ``critical'' features of the statistical properties of earthquakes lead to the view that earthquake might be a phenomenon of ``self-organized criticality (SOC)''. This view is opposite to the other widely-spread view of ``characteristic earthquakes'' where an earthquake is regarded to possess its characteristic energy or time scale. 

 Modeling earthquakes and elucidating its statistical properties have been a fruitful strategy in earthquake studies.  One of the popular model might be the so-called spring-block model originally proposed by Burridge and Knopoff (BK) [\textit{Burridge and Knopoff}, 1967]. In this model, an earthquake fault is simulated by an assembly of blocks, each of which is connected via the elastic springs to the neighboring blocks and to the moving plate. All blocks are subject to the friction force, the source of the nonlinearity in the model, which eventually realizes an earthquake-like frictional instability. While the spring-block model is obviously a crude model to represent a real earthquake fault, its simplicity enables one to study its statistical properties with high precision.

 Carlson, Langer and others [\textit{Carlson and Langer}, 1989a; \textit{Carlson and Langer}, 1989b; \textit{Carlson et al.}, 1991; \textit{Carlson}, 1991a; \textit{Carlson}, 1991b; \textit{Carlson et al.}, 1994]  studied the statistical properties of the BK model quite extensively, paying particular attention to the magnitude distribution of earthquake events. Most of these simulations have been done for the simplest one-dimensional (1D) version of the model. It was then observed that, while smaller events persistently obeyed the GR law, {\it i.e.\/}, staying critical or near-critical, larger events exhibited a significant deviation from the GR law, being off-critical or ``characteristic''[\textit{Carlson and Langer}, 1989a; \textit{Carlson and Langer}, 1989b; \textit{Carlson et al.}, 1991; \textit{Carlson}, 1991a; \textit{Carlson}, 1991b; \textit{Schmittbuhl et al.}, 1996].  The spring-block model has also been extended in several ways, {\it e.g.\/}, taking account of the effect of the viscosity [\textit{Myers and Langer}, 1993; \textit{Shaw}, 1994; \textit{De and Ananthakrisna}, 2004], modifying the form of the friction force [\textit{Myers and Langer}, 1993; \textit{Shaw}, 1995; \textit{Cartwright}, 1997; \textit{De and Ananthakrisna}, 2004], considering the long-range interactions between blocks [\textit{Xia et al.}, 2005, 2007], driving the system only at one end of the system [\textit{Vieira}, 1992], or by incorporating the rate- and state-dependent friction law [\textit{Ohmura and Kawamura}, 2007]. 

 In the previous paper, the present authors studied the statistical properties of the 1D BK model, focusing on its {\it spatiotemporal correlations\/} [\textit{Mori and Kawamura}, 2005;2006]. This study has revealed several interesting features of the 1D BK model. For example, preceding the mainshock, the  frequency of smaller events is gradually enhanced, whereas, just before the mainshock, it is suppressed in a close vicinity of the epicenter of the upcoming event (the Mogi doughnut). The time scale of the onset of the doughnut-like quiescence depends on the extent of the frictional instability. Furthermore, the apparent $B$-value of the magnitude distribution increases significantly preceding the mainshock under certain conditions. 

 It should be remembered, however, that real earthquake faults are 2D rather than  1D. Hence, it is clearly desirable to study the 2D version of the model in order to further clarify the statistical properties of earthquakes. In the present paper, we  study spatiotemporal correlations of the 2D BK model. The 2D model studied here is intended to represent a 2D fault plane itself, where the direction orthogonal to the fault plane is regarded to be rigid and not considered explicitly in the model [\textit{Carlson}, 1991b]. The other possible version of the 2D model is the one where the second direction of the model is taken to be orthogonal to the fault plane [\textit{Myers et al.}, 1996]. 

 Previous calculations on the 2D BK model were performed with main interest in their magnitude distribution, while there were very few systematic studies of its spatiotemporal correlations. In the present paper, we wish to fill this gap by investigating the spatiotemporal correlations of the 2D BK model, extending our previous study on the 1D BK model [\textit{Mori and Kawamura}, 2005; 2006]. Statistical properties of the model are examined here for quite a wide range of the model parameters. Concerning the most relevant parameter representing the extent of the frictional instability $\alpha$, we study the properties of the model with varying the parameter $\alpha$ over all possible range of [$0,\infty$]. Such comprehensive studies of the 2D BK model enables us to identify several regions of the parameter space characterized by qualitatively different behaviors.

 The present paper is organized as follows. In \S 2, we introduce the model
and explain some of the details of our numerical simulation. 
The results of our simulations are presented in \S 3 and \S 4. In \S 3, 
we show the results of the event-size distribution, and construct a ``phase diagram'' in the parameter space characterizing the model. The properties of the mean displacement and the mean number of failed-blocks of each seismic event are analyzed. The possible effect of the spatial anisotropy of the elastic parameters is also examined. Then,  in section 4, we analyze
various types of spatiotemporal correlation functions of seismic events, including the local recurrence-time distribution, the time-correlation function of seismic events before and after the mainshock, the time development of the spatial correlation function of seismic events before and after the mainshock, and the time development of the magnitude distribution function  before and after the mainshock.
Finally, \S 5 is devoted to summary and discussion.

%
%


%
%

\section{The model and the method}

Our model is the 2D version of the spring-block BK model, representing a ``fault plane'', which is taken to be the $x-z$ plane.
The plane consists of a 2D square array of blocks containing
$N_x$ blocks in the $x$-direction and $N_z$ blocks in the $z$-direction. 
All Blocks are assumed to move only in the $x$-direction along strike, and 
are subject to the friction force $\Phi$.
Each block is connected with its four nearest-neighbor blocks
 via the springs of the elastic constant $k_c$,
 and is also connected to the moving plate via the spring
 of the elastic constant $k_p$. For simplicity, we assume for most part 
that the spring constants are spatially isotropic, {\it i.e.\/}, assume that the elastic constant of the spring in the $x$-direction $k_{cx}$ and the one in the $z$-direction $k_{cz}$ are identical, $k_{cx}=k_{cz}=k_c$. (Later, the effect of the spatial anisotropy of the spring constants $k_{cx}\neq k_{cz}$ will be studied briefly.)

The equation of motion for the block at the site $(i,j)$ is given by
\begin{equation}
\begin{array}{ll}
m \ddot U_{i,j}=k_p (\nu ' t'-U_{i,j}) + k_c (U_{i+1,j}+U_{i,j+1} \ \ \
 \ \
\\ \ \ \ \ \ \ +U_{i-1,j}+U_{i,j-1}-4U_{i,j})-\Phi (\dot U_{i,j}),
\end{array}
\end{equation}
where $m$ is the mass of a block, $t'$ is the time, $U_{i,j}$ is the displacement  along the $x$-direction of the  block $(i,j)$, and $\nu '$ is the loading rate 
representing the speed of the plate.

The equation is made dimensionless in the same way as in [\textit{Mori and Kawamura}, 2006], {\it i.e.\/}, 
the time $t'$ is measured in units of the characteristic 
frequency $\omega =\sqrt{k_p/m}$ and the displacement $U_{i,j}$ in units of
the length $L=\Phi(0)/k_p$, $\Phi(0)$ being a static friction. Then,
the equation of motion can be written  in the dimensionless form as
\begin{equation}
\begin{array}{ll}
\ddot u_i=\nu t-u_{i,j}+l^2(u_{i+1,j}+u_{i,j+1} \ \ \ \ \ \\
\ \ \ \ \  +u_{i-1,j}+u_{i,j-1}-4u_{i,j})-\phi (\dot u_i),
\end{array}
\end{equation}
where $t=t'\omega $ is the dimensionless time, 
$u_{i,j}\equiv U_{i,j}/L$ is the dimensionless displacement of the 
block ($i,j$), 
$l \equiv \sqrt{k_c/k_p}$ is the dimensionless stiffness parameter, 
$\nu =\nu '/(L\omega)$ is the dimensionless loading rate, and  
$\phi(\dot u_i) \equiv \Phi(\dot U_i)/\Phi(0)$ is the dimensionless friction 
force.

As the form of the friction force $\phi(\dot u_i)$, 
we use the velocity-weakening friction force which is a single-valued function of the velocity. As its explicit functional form, we 
use the form introduced by [\textit{Carlson et al.}, 1991],
\begin{equation}
\phi(\dot u) = \left\{ 
             \begin{array}{ll} 
             (-\infty, 1],  & \ \ \ \ {\rm for}\ \  \dot u\leq 0, \\ 
              \frac{1-\sigma}{1+2\alpha \dot u/(1-\sigma )}, &
             \ \ \ \ {\rm for}\ \  \dot u>0, 
             \end{array}
\right.
\end{equation} 
where the friction force immediately drops to $1-\sigma$ on sliding, and decays toward zero with a rate proportional to the parameter $\alpha$ as the velocity increases. 
%
%
The back-slip is inhibited by imposing an
infinitely large friction for $\dot u_i<0$, {\it i.e.\/}, 
$\phi(\dot u<0)=-\infty $. 

This friction force represents the {\it velocity-weakening\/} friction force.
Real friction force is of course more complex, not depending on 
the velocity alone. Indeed, various types of constitutive laws describing rock friction have been proposed in the literature, {\it e.g.} 
the slip-weakening friction 
force [\textit{Scholz}, 2002;\textit{Shaw},
1995;\textit{Myers et al.}, 1996;\textit{Scholz}, 1998] 
or the rate- and state-dependent 
friction force [\textit{Dietrich}, 1979;\textit{Ruina}, 1983;\textit{Scholz}, 2002;\textit{Scholz}, 1998], {\it etc\/}. Here, in order to investigate the effect of the dimensionality on the spatiotemporal correlations of the BK model, we assume the simplest version of the velocity-weakening friction force.

The friction force is characterized by the two 
parameters, $\sigma$ and $\alpha$. 
The former, $\sigma$, 
represents an instantaneous drop of the friction force
at the onset of the slip, while the latter, $\alpha$, 
represents the rate of the friction force getting weaker
on increasing the sliding velocity. The $\alpha = 0$ case represents the simplest Coulomb friction law  where the friction force instantaneously drops from the static value $1$ to its dynamical value $1-\sigma$  as soon as the block begins to slide, and is kept constant on sliding irrespective of the velocity. The $\alpha =\infty$ case also corresponds to the another Coulomb friction law where the dynamical friction immediately drops to zero on sliding.  In addition to these frictional-parameters, the model possesses one more material parameter, an elastic-parameter $l$. We try to cover a rather wide range of the parameters $\alpha$, $l$ and $\sigma$. For example, we take the frictional-parameter $\alpha$ in the range $\alpha=[0, \infty]$, and systematically examine the $\alpha$-dependence of the results.

 We also assume the loading rate $\nu$ to be infinitesimally small, and put 
$\nu=0$ during an earthquake event, a very good approximation 
for real faults [\textit{Carlson et al.}, 1991]. Taking this limit ensures that the interval time during successive earthquake events can be measured in units of $\nu^{-1}$ irrespective of particular values of $\nu$. 

 Numerical details are the same as in [\textit{Mori and Kawamura}, 2006].
We solve the equation of motion (2) 
by using the Runge-Kutta method of the fourth 
order, the width of the time discretization $\Delta t$ being
$\Delta t\nu =10^{-6}$. Total number of $10^5 \sim 10^7$  events
are generated in each run, 
which are used to perform various averagings. In calculating the observables,
initial $10^5$ events are discarded as transients.

 When the model is regarded to represent a strike-slip fault, the $z$-direction is supposed to correspond to the depth direction. Following [\textit{Carlson}, 1991b], we impose periodic boundary condition in the $x$-direction and free boundary condition in the $z$-direction. 
For the most part of our calculation, the system size is taken to be $N_x=160$ and $N_z=80$. In some cases, we study several other sizes, $N_x=N_z=N$, with varying $N$ in the range $60\leq N\leq 480$.  


%
%

\section{Event-size distribution}

In this and following sections, we show the results of our numerical simulations on the 2D BK model for various observables. In this section, we analyze the event-size distributions of earthquakes.

\subsection{Magnitude distribution}


We define the magnitude of an event, $\mu$, as a logarithm of its moment $M$, {\it i.e.\/}, 
\begin{equation}
\mu= \ln M = \ln \left( \sum_{i,j} \Delta u_{i,j} \right),
\end{equation}
where $\Delta u_{i,j}$ is the total displacement of the {$i,j$} block during a given event and the sum is taken over all blocks involved in the event [\textit{Carlson et al.}, 1991]. 

 Typical behaviors of the magnitude distribution of the 2D BK model are shown in Figs.1(a) and (b) for the case of $l=3$ and $\sigma=0.01$, with varying the frictional-parameter $\alpha$. The magnitude distribution $R(\mu){\rm d}\mu$ represents the rate of events with their magnitudes in the range [$\mu, \mu + {\rm d}\mu$]. The parameter $\alpha$ is varied in a wide range of $0\leq \alpha \leq \infty$.
 
  Fig.1(a) exhibits $R(\mu)$ for smaller $\alpha$. In the range
  $\alpha \lsim 0.5$, $R(\mu)$ bends down rapidly at larger
  magnitudes. Such a behavior of $R(\mu)$ is often called
  ``subcritical''. Only small events of $\mu \lsim 2$ occur in this
  case.   For $\alpha \gsim 0.5$, large earthquakes of their
 magnitudes $6 \lsim \mu \lsim 8$  suddenly appear, while earthquakes
of intermediate magnitudes, say, $2\lsim \mu \lsim 6$, remain rather
scarce. It means that large and small earthquakes are well separated
at $\alpha\simeq 0.5$. Such a sudden appearance of large earthquakes
at $\alpha =\alpha_{c1}\simeq 0.5$ coexisting with smaller ones has a
feature of a ``discontinuous transition''. Indeed, the maximum magnitude of the observed events (within our observation time) $\mu_{max}$ suddenly jumps from  $\mu_{max}\simeq 4$ at $\alpha=0.4$ to $\mu_{max}\simeq 7$ at $\alpha=0.5$, demonstrating the discontinous nature of the transition at $\alpha\simeq 0.5$. In this connection, it should
be mentioned that Vasconcelos showed that a single-block system
exhibited a first-order 
transition at $\alpha =1$ from a stick-slip to a creep
[\textit{Vasconcelos}, 1996], whereas this first-order transition
becomes apparently continuous in the 1D many-block system [\textit{Vieira
et al.}, 1993; \textit{Clancy and Corcoran}, 2005]. A discontinuous change 
observed at $\alpha=\alpha_{c1}\simeq 0.5$ in the present
2D model may be related to the first-order transition of a
single-block system, although events observed at  $\alpha<\alpha_{c1}$
in the present 2D model are not really creeps, but rather are
stick-slip events of small sizes. 

 With increasing $\alpha$ further, earthquakes of intermediate
 magnitudes gradually increase their frequency.  Fig.1(b) exhibits
 $R(\mu)$ for larger $\alpha$. In the range of $1\lsim \alpha \lsim
 10$,  $R(\mu)$ exhibits a pronounced peak structure at a larger
 magnitude, deviating from the GR law at  $\mu\gsim \tilde \mu\simeq 2$, 
while it exhibits a near straight-line behavior  corresponding to the GR law at smaller magnitudes $\mu\lsim \tilde \mu$. The observed behavior of $R(\mu)$ agrees with the previous findings by Carlson, Langer and collaborators [\textit{Carlson and Langer},1989a; \textit{Carlson and Langer} 1989b; \textit{Carlson et al.}, 1991]. Such a  behavior of $R(\mu)$ is sometimes called ``supercritical'', since $R(\mu)$ bends up at larger magnitudes (though it eventually falls off at still larger magnitudes). According to Carlson {\it et al.}, smaller events of their magnitudes $\mu\lsim \tilde \mu$ are localized events, while larger events of their magnitudes $\mu\gsim \tilde \mu$ are delocalized ones. The existence of a distinct peak structure at a larger magnitude suggests that large earthquakes of $\mu\gsim \tilde \mu$ are more or less characteristic. In order to check that the peak structure in $R(\mu)$ is not an artifact due to the finite-size effect, the system-size dependence of $R(\mu)$ is examined in Fig.1(c) for the case of $\alpha=3$, for $N\times N$ systems with $60\leq N\leq 480$. As can be seen from the figure, the characteristic peak at a larger magnitude is affected somewhat by the finite-size for $N\lsim 240$, while the peak  is almost $N$-independent for $N\gsim 240$. This observation indicates that the characteristic peak structure of $R(\mu)$ is an intrinsic property of the bulk system.
 
 As $\alpha$ increases further, the peak  at a larger magnitude becomes less pronounced.  At $\alpha =\alpha_{c2}\simeq 13$, $R(\mu)$ exhibits a near straight-line behavior for a rather wide magnitude range, though $R(\mu)$ falls off rapidly at still larger magnitudes $\mu \gsim 7$, indicating that the ``near-critical'' behavior observed for $\alpha=\alpha_{c2}\simeq 13$ cannot be regarded as a truly asymptotic one. We have checked that this rapid fall-off of $R(\mu)$ at very large magnitudes is  a bulk property, not a finite-size effect: See the inset of Fig.1(c).
 On further increasing $\alpha$ beyond $\alpha >\alpha_{c2}\simeq 13$,
 $R(\mu)$ exhibits again a bending-down ``subcritical'' behavior. The
 change from the ``supercritical'' to ``subcritical'' behavior at
 $\alpha=\alpha_{c2}\simeq 13$ is a continuous one, in
 contrast to the discontinuous change observed at $\alpha=\alpha_{c1}\simeq 0.5$. 

 The behavior of $R(\mu)$ remains qualitatively the same even when the value of $l$ is varied. In Figs. 2 and 3, we show $R(\mu)$ both for a smaller value of $l$, $l=1$, and for a larger value of $l$, $l=5$, respectively. In either case, the behavior of $R(\mu)$ is qualitatively similar to the one shown in Figs.1 for the case of $l=3$. In particular, on increasing $\alpha$, the ``subcritical'' behavior realized at smaller $\alpha$ changes into the ``supercritical'' behavior via a ``discontinuous transition'' at  $\alpha=\alpha_{c1}$, and then, into the ``subcritical'' behavior realized at larger $\alpha$ via a ``continuous transition'' at  $\alpha=\alpha_{c2}$. 

 In Fig.4, we summarize the behavior of $R(\mu)$ in the form of a ``phase diagram'' in the frictional-parameter $\alpha$ versus the elastic-parameter $l$ plane for the case of $\sigma=0.01$. The phase diagram consists of three distinct regions,  two of which are  ``subcritical'' regions and one is ``supercritical'' region. The ``phase boundary'' between the smaller-$\alpha$  ``subcritical'' region and the ``supercritical'' region represents a ``discontinuous transition'', while the one between the larger-$\alpha$ ``subcritical'' region and the ``supercritical'' region represents a ``continuous transition''. Note that the ``phase boundary'' is given here by a visual inspection of the magnitude distribution $R(\mu)$ and of several other quantities given in \S 3.3 below. No detailed scaling analysis as performed in [\textit{Clancy and Corcoran}, 2005] has been made. The transition between these different ``phases'', {\it i.e.\/}, a ``subcritical phase'' for small $\alpha$, a ``supercritical phase'' for intermediate $\alpha$, and another ``subcritical phase'' for large $\alpha$, is primarily dictated by the $\alpha$-value. Since the ``continuous transition'' in Fig.4 has a finite slope in the $\alpha$-$l$ plane,  one can also induce the ``subcritical''-``supercritical'' transition by increasing the $l$-value for a fixed $\alpha$. A similar  transition induced by the change in the $l$-value was reported also for the 1D BK model [\textit{Espanol}, 1994;\textit{Vieira}, 1996]

 So far, we have fixed $\sigma$ to $\sigma=0.01$.  Next, we analyze the $\sigma$-dependence of the magnitude distribution. In Figs.5, we show $R(\mu)$ for various values of $\alpha$ both for a larger value of $\sigma$, $\sigma=0.1$, and  for a smaller value of $\sigma$, $\sigma=0.001$ with fixing $l=3$. As can be seen from the figures, qualitative features of $R(\mu)$ remain essentially the same even including its $\alpha$-dependence as in the case of $\sigma=0.01$. In particular, on increasing $\alpha$, the system exhibits a discontinuous change from the small-$\alpha$ ``subcritical'' behavior to the intermediate-$\alpha$ ``supercritical'' behavior at $\alpha=\alpha_{c1}$, then a continuous change from  the ``supercritical'' behavior to the large-$\alpha$ ``subcritical'' behavior at $\alpha=\alpha_{c2}$. Hence, the $\alpha$-$l$ phase diagrams for $\sigma=0.1$ and $\sigma=0.001$ are qualitatively similar to the one of $\sigma=0.01$ shown in Fig.4. Concerning the values of $\alpha_{c1}$ and $\alpha_{c2}$, $\alpha_{c1}\simeq 0.5$ is insensitive to the $\sigma$-value, while $\alpha_{c2}$ tends to decrease with increasing $\sigma$. Hence,  on increasing $\sigma$, the width of the ``supercritical'' region is gradually narrowed. In fact, we have observed that the ``supercritical'' region vanishes for $l=3$ at around $\sigma\simeq 0.5$. Hence, for sufficiently large $\sigma$ close to $\sigma=1$, the ``supercritical'' region is expected to vanish in the $\alpha$-$l$ phase diagram, only a single ``subcritical'' region being left.

 In Fig.6, we show $R(\mu)$ for various values of $\sigma$ with fixing $\alpha=0$.   Note that this $\alpha=0$ case corresponds to the case of a constant dynamic friction of strength $1-\sigma$. As can be seen from Fig.5, $R(\mu)$ always exhibits the ``subcritical'' behavior for all values of $\sigma$. Namely, the constant dynamical friction always gives rise to the ``subcritical'' behavior.  In particular, the case ($\alpha=0$, $\sigma=1$) corresponding to the vanishing dynamical friction is equivalent to the case ($\alpha=\infty$, $\sigma$). This is fully consistent with our observation that the $\alpha=\infty$ limit always yields a ``subcritical'' behavior in Figs.1-3 and 5.

\subsection{Effect of spatial anisotropy}

 So far, we have assumed for simplicity that the elastic-parameter $l$ is spatially isotropic within the 2D plane, {\it i.e.\/}, have assigned the same $l$-values to the spring constants in the $x$- and in the $z$-directions,  $l_x=l_z=l$. Kumagai {\it et al\/} suggested that, in order to better mimic the original continuum crust, the spring constants of the BK model should be set spatially anisotropic, {\it i.e.\/}, $l_x=\sqrt 3 l_z$ [\textit{Kumagai et al.}, 1999]. Thus, we also analyze in this subsection the effect of spatial anisotropy of the elastic-parameter $l$ on $R(\mu)$. 

 In Fig.7, we show $R(\mu)$ for the case of $\alpha$=3.5, $l_x=\sqrt 3$, $l_z=1$ and $\sigma =0.01$, which correspond to exactly the same parameter values as used in [\textit{Kumagai et al.}, 1999]. In the same figure, we also show $R(\mu)$ of the corresponding isotropic system, $l=l_x=l_z=(1+\sqrt 3)/2$ (all other parameter are chosen common) for comparison. This value of $l$ is chosen to be a mean of the anisotropic $l$-values given above. Note that the point ($\alpha=3.5$, $l=(1+\sqrt 3)/2,\ \sigma=0.01$) lies close to the ``continuous'' phase boundary in Fig.4. As can be seen from Fig.7, $R(\mu)$ of the isotropic system turns out to be quite similar to that of the anisotropic system. We thus conclude that the spatial anisotropy of the elastic parameters does not cause any qualitative new feature to the properties of the corresponding spatially isotropic model.

\subsection{The mean displacement and the mean number of failed-blocks}

 The size of an earthquake event is usually measured by its magnitude. Other possible measures of event size might be the mean displacement $\Delta \bar u$, the number of failed-blocks $N_b$ (corresponding to the size of rupture zone), and the mean stress-drop.  Note that, in the BK model, the mean stress-drop of an event is essentially identical with (proportional to) the mean displacement of an event. In Figs.8 and 9, we respectively show the mean displacement and the mean number of failed-blocks as a function of the magnitude for various values of $\alpha$. Naturally, all the quantities are increasing functions of the magnitude. An interesting observation here is that the data in the ``subcritical'' regimes are grouped into two distinct branches both in Figs.8(a) and 9(a), each corresponding to the small-$\alpha$ and large-$\alpha$ ``subcritical'' regions of Fig.4.

 As can be seen from Fig.8(a), the data in the small-$\alpha$ ``subcritical'' regime ($\alpha < \alpha_{c1} \simeq 0.5$) lacks events of larger magnitudes and are characterized by smaller displacement, while those in the large-$\alpha$ ``subcritical'' regime ($\alpha > \alpha_{c2} \simeq 13$) are characterized by much larger displacement, by about a factor $10^2\approx \sigma^{-1}$ larger than that in the small-$\alpha$ ``subcritical'' regime with the same $\mu$. All the data of the mean displacement $\Delta \bar u$ in the ``subcritical'' regimes collapse, at least approximately,  onto these two curves, which are both linear in the magnitude with a common slope $\simeq 0.1$. By contrast, the data in the ``supercritical'' regime ($\alpha_{c1} < \alpha < \alpha_{c2}$) exhibit a significantly different behavior. At smaller magnitudes $\mu\lsim  5$, they exhibit a crossover behavior between these two universal ``subcritical'' curves depending on its $\alpha$-value: For smaller $\alpha$ close to $\alpha_{c1}$, the data tend to lie closer to the small-$\alpha$ ``subcritical'' curve, while for larger $\alpha$ close to $\alpha_{c2}$, the data tend to lie closer to the large-$\alpha$ ``subcritical'' curve. At larger magnitudes $\mu\gsim 5$, the data in the ``supercritical'' regime seem to approach the asymptotic straight line again. At larger magnitudes, however, finite-size effects are significant. Hence, we show in Fig.8(b) the system-size dependence of the mean displacement at larger magnitudes, for $N\times N$ systems with $N$ in the range $60\leq N\leq 480$. As can be seen from the figure, the ``supercritical'' curve at large magnitudes appear to approach a straight line with a slope $\simeq 0.1$, which, however, is not a simple continuation of the large-$\alpha$ ``subcritical'' curve. 

 The existence of the two ``subcritical'' curves and the asymptotic ``supercritical'' curve at $\mu\gsim 5$  is also clearly visible in Figs.9 for the magnitude dependence of the mean number of failed-blocks $\bar N_b$. The two ``subcritical'' curves are again both strikingly linear with a common slope $\simeq 0.9$, while the asymptotic ``supercritical'' curve at $\mu\gsim 5$ is also apparently linear with a slope $\simeq 0.9$. 

 In order to get further insights into the nature of the ``subcritical'' behavior, we show in Figs.10(a) and (b) the magnitude dependence of the mean displacement and of the mean number of failed-blocks for various values of $\sigma$, $\sigma=0.001, 0.01, 0.1, 1$, for the case of $\alpha=0$ which corresponds to a constant dynamical friction of strength $1-\sigma$. As already shown in Fig.6, all these $\alpha=0$ cases correspond to the ``subcritical'' regime. As expected, all the data of the mean displacement show linear behaviors with a common slope $\simeq 0.1$, while those of the mean number of failed-blocks show linear behaviors with a common slope $\simeq 0.9$. An interesting observation here is that the data for different values of $\sigma$ are equally spaced both in Figs.10(a) and (b), meaning that, in the case of a constant dynamical friction, the mean displacement (the mean number of failed-blocks) is proportional to the parameter $\sigma$ ($\sigma^{-1}$). 

 Hence, the 2D BK model exhibits three distinct behaviors depending on the parameters, {\it i.e.\/}, the small-$\alpha$ ``subcritical'' behavior at $\alpha<\alpha_{c1}$, the ``supercritical'' behavior at $\alpha_{c1} < \alpha < \alpha_{c2}$ and the large-$\alpha$ ``subcritical'' behavior at $\alpha>\alpha_{c2}$.  Given its magnitude, the small-$\alpha$ ``subcritical'' behavior is characterized by smaller displacement, smaller stress-drop and larger rupture-zone size (the number of failed-blocks) with more weight on smaller magnitudes, while the large-$\alpha$ ``subcritical'' behavior is characterized by larger displacement, larger stress-drop and smaller rupture-zone size with more weight on larger magnitudes. Seismic events in the ``subcritical'' regime exhibit an interesting scaling property, {\it i.e.\/}, $\Delta \bar u\propto M^{0.1}$ and $\bar N_b \propto M^{0.9}$, $M\equiv \exp \mu$ being a seismic moment of an event. Seismic events in the ``supercritical'' regime exhibit, at smaller magnitudes, a crossover behavior between the two ``subcritical'' behaviors, and exhibit the scaling property at larger magnitudes.



\section{Spatiotemporal correlations of earthquakes}

\subsection{Local recurrence-time distribution}

 Earthquake recurrence, {\it i.e.\/}, how earthquakes repeat in time, is a question of general interest, and is closely related to the issue of criticality versus periodicity. Characteristic earthquake recurrence would mean the existence of a characteristic time scale for earthquake recurrence, while critical earthquake recurrence would mean the absence of any such characteristic time scale. We investigate earthquake recurrence of the 2D BK model via the {\it local\/} recurrence-time distribution function.  The local recurrence time $T$ is defined by the time passed until the next event occurs with its epicenter lying in a vicinity of the previous event within distance of $r$-blocks from the epicenter of the previous event. Here, we consider events  with their magnitude $\mu \geq \mu _c$ ($\mu_c=5,3,0$), and compute the distribution of the local recurrence time $T$  with $r=5$. 

 In the main panels of Figs.11(a) and (b), we show on a log-log plot the computed local distribution function $P(T)$ for the case of $\alpha=3$ (a), and $\alpha=20$ (b), with fixing $l=3$ and $\sigma=0.01$, each corresponding to the ``supercritical'' and ``subcritical'' regimes, respectively. In the insets, the same data are re-plotted on a semi-logarithmic scale. The recurrence time is normalized by its mean $\bar T$, which is $\bar T\nu =14.0$, 19.5, 22.5, respectively for $\mu_c=0$, 3, 5 in the case of Fig.11(a), and  $\bar T\nu =0.14$, 1.38, 35.5, respectively for $\mu_c=0$, 3, 5 in the case of Fig.11(b).

As can be seen from the main panels of the figure, $P(T)$ exhibits
different behaviors between in the ``supercritical'' and
``subcritical'' regimes: In the ``supercritical'' regime, as shown in
Fig.11(a), $P(T)$ exhibits two peaks (or a peak and a shoulder), 
one at a shorter time $T/\bar T \sim 10^{-2}$ and the other at a longer time $T/\bar T \sim 10^{-1}$, whereas in the ``subcritical'' regime, as shown in Fig.11(b), $P(T)$ with $\mu_c=5$ exhibits a single peak only. In either case, $P(T)$ exhibits an exponential tail at longer times: See the insets of Figs.11. Similar exponential tail is also observed in real seismicity [\textit{Corral}, 2004].

 The existence of peak structures and an exponential tail in $P(T)$  means the existence of characteristic time scales in earthquake recurrence in the 2D BK model. We note that similar characteristic behavior of $P(T)$ was also observed in the corresponding 1D BK model [\textit{Mori and Kawamura}, 2005; 2006].

\subsection{Time correlation of events associated with the mainshock}

In real seismicity, large events often accompany foreshocks and aftershocks. In Fig.12, we show the time correlation function between large events (mainshock) and events of arbitrary sizes (dominated in number by small events) for various values of the frictional-parameter $\alpha$ with fixing $l=3$ and $\sigma=0.01$. In the figure, we plot the mean number of events of arbitrary sizes occurring within 5 blocks from the epicenter of the mainshock before  ($t<0$) and after ($t>0$) the mainshock, where the occurrence of the mainshock is taken to be the origin of the time $t=0$. The average is taken over all large events of their magnitudes with $\mu \geq \mu_c=5$. The number of events are counted here with the time bin of $\Delta t\nu =0.02$.

 In the ``supercritical'' regime, as can be seen from Fig.12,
 a remarkable acceleration of seismic activity occurs  before the
 mainshock ($t<0$). 
This feature is also observed 
in the corresponding 1D BK model [\textit{Mori and Kawamura}, 2006].  For the case of $\alpha =3$ lying in the midst of the ``supercritical'' regime,   seismic activity stays low after the mainshock ($t>0$), and is gradually enhanced as the time passes. This calm period continues during about half of the recurrence time. After this calm period, the seismic activity is activated. For the cases of $\alpha =1$ and 10  near the border of the ``subcritical'' regime,   seismic activity after the mainshock still remains relatively high and is gradually suppressed as the time passes. We note that such strong time correlations observed in the ``supercritical'' regime are robust against the change of the magnitude of the mainshock. In particular, even if one chooses as a mainshock an event of a smaller magnitude-threshold $\mu_c < 5$, one still gets eminent time correlations quite similar to the ones shown in Fig.12 as long as one is in the ``supercritical'' regime.

 In the ``subcritical'' regime, by contrast, the time correlation is almost absent except for the suppression of seismicity immediately before the mainshock, as can be seen from Fig.12.

\subsection{Spatial correlations of events before the mainshock}

 In this subsection, we examine the time-development of spatial correlations of seismic events {\it before\/} the mainshock. In Figs.13(a) and (b), we show the spatial correlation functions between the mainshock and the preceding events of arbitrary size, dominated in number by small events, for several time periods  before the mainshock, for the cases of $\alpha=3$ (a) and $\alpha=20$ (b) with fixing $l=3$ and $\sigma=0.01$, each corresponding to the ``supercritical'' and ``subcritical'' regimes, respectively. It represents the conditional probability that, provided that a large event with $\mu >\mu_c =5$ occurs at a time $t_0$, an event of arbitrary size occurs at a time $t-t_0 < 0$ with its epicenter lying within distance $r$ from the epicenter of the mainshock. The computed spatial correlation functions are shown as a function of $r$. Insets represent longer-time behaviors. 

 In the ``supercritical'' regime,  as shown in Fig.13(a), the frequency of small events are enhanced preceding the mainshock at and around the epicenter of the upcoming mainshock. For small enough $t$, such a cluster of smaller events correlated with the large event may be regarded as foreshocks. The eminent spatial correlations observed in the ``supercritical'' regime turns out to be robust against the change of the magnitude of the mainshock. Just before the mainshock, the frequency of smaller events is suppressed in a close vicinity of the upcoming mainshock, while it continues to be enhanced in the surrounding blocks. The spatial range where the quiescence occurs is narrow, only of a few blocks. Such a doughnut-like quiescence phenomenon prior to the mainshock was observed also in the 1D BK model. In real seismicity, the doughnut-like quiescence has been known as the ``Mogi doughnut'' [\textit{Mogi}, 1969;\textit{Mogi}, 1979;\textit{Scholz}, 2002]. 

 In the ``subcritical'' regime,  as shown in Fig.13(b), the seismic acceleration preceding the mainshock is hardly discernible, while the doughnut-like quiescence is still realized. The observation is consistent with the result of the time correlations shown in Fig.12.

 The doughnut-like quiescence is observed both in the ``supercritical'' and ``subcritical'' regimes in common. The time scale for the appearance of the doughnut-like quiescence depends on the  $\alpha$- and $l$-values.  Namely, the time scale of the onset of the doughnut-like quiescence tends to be shorter for smaller $\alpha$ or for larger $l$. The spatial range where the quiescence occurs also weakly depends on the parameter $\alpha$, getting longer for larger $\alpha$.  As mentioned, the observed quiescence is quite similar to the one observed in the 1D BK model. In the case of the 1D BK model, the doughnut-like quiescence can be understood in terms of characteristic interval-time scales associated with a single-block event [\textit{Mori and Kawamura}, 2006]. In fact, we have confirmed that essentially the same mechanism operates even in the present 2D model, {\it i.e.\/}, single-block events give rise to the observed doughnut-like quiescence phenomena.

\subsection{Spatial correlation of events after the mainshock}

 In this subsection, we examine the time-development of spatial correlations of seismic events after the mainshock. The calculated spatial correlation functions are shown in Figs.14(a)-(d), for the cases of the ``supercritical'' regime, $\alpha=1$ (a), $\alpha=3$ (b), $\alpha=10$ (c), and in the ``subcritical'' regimes, $\alpha=20$ (d). Other conditions are taken to be the same as in Figs.13. Insets represent shorter-time behaviors. 

 As can be seen from these figures, spatial correlations are eminent in the ``supercritical'' regime (Figs.14(a)-(c)), but are almost absent in the ``subcritical'' regime (Fig.14(d)). Details of the spatial correlations observed in the ``supercritical'' regime, however, depends on the parameter values somewhat. As can be seen from Figs.14(a) and (c), seismic events for $\alpha =1$ and 10 remain active in the vicinity of the epicenter of the mainshock, which are gradually suppressed as the times passes. One may regard this as ``aftershock''. However, as can be seen from the insets, at shorter time scales, the frequency of events stays almost constant in time. Furthermore, even in the time range where seismic activity decays monotonically in time, its time-dependence does not obey the Omori law. 

 Somewhat different behavior is observed for the case of $\alpha=3$ lying in the midst of the ``supercritical'' regime. As can be seen from Fig.14(b), seismic activity is rather suppressed in a vicinity of the epicenter of the mainshock just after the mainshock, staying almost constant in time for some period, and is eventually gradually enhanced. Again, the eminent spatial correlations observed in the ``supercritical'' regime is robust against the change of the magnitude of the mainshock.

 In any case, aftershocks obeying the Omori-law generically observed in real seismicity is not realized in the 2D BK model for any parameter value. This feature is also common to the 1D BK model [\textit{Carlson and Langer},
1989a;\textit{Carlson and Langer}, 1989b;\textit{Mori and Kawamura}, 2006].

\subsection{Time-dependent magnitude distribution}

 In real seismicity, an appreciable change of the $B$-value of the magnitude distribution has been reported preceding large earthquakes: Often a decrease of the $B$-value [\textit{Suyehiro et al.}, 1964; \textit{Jaume and Sykes}, 1999; \textit{Kawamura}, 2006], but sometimes an increase of it [\textit{Smith}, 1981]. Obviously, a possible change in the magnitude distribution preceding the mainshock possesses a potential importance in earthquake prediction.

 In Figs.15, we show  the ``time-resolved''  local magnitude distributions for several time periods {\it before\/} the large event for the cases of $\alpha =1$ (a), $\alpha=13$ (b) and $\alpha =20$ (c) with fixing $l=3$ and $\sigma=0.01$, each corresponding to the ``supercritical'', ``near-critical'' and ``subcritical'' regimes, respectively. Only events with their epicenters lying within 5 blocks from the upcoming mainshock is counted here. We define the mainshock as a large event of $\mu\geq \mu_c=5$. In the ``supercritical'' case, as can be seen from Fig.15(a),  an apparent $B$-value describing the smaller magnitude region, $\mu\lsim 2$, gets smaller as the mainshock is approached, {\it i.e.\/}, it changes from $B\simeq 0.89$ of the long-time value to  $B\simeq 0.65$ in the time range $t\nu \leq 0.1$ before the mainshock. In the ``subcritical'' regime, on the other hand, an apparent $B$-value gets larger as the mainshock is approached, as can be seen from Fig.15(c). In the ``near-critical'' case, an apparent $B$-value hardly changes even when the mainshock is approached, although the magnitude distribution loses its weight at larger magnitudes: See Fig.15(b). 


 In real seismicity, the $B$-value usually decreases just before the mainshock, although the opposite tendency, {\it i.e.\/}, an increase of the $B$-value,  was also reported. Our present result might give a hint to understand such a complex change of the $B$-value  observed in real earthquakes. The analysis of the JUNEC seismic catalogue, covering earthquakes in Japan area during the period 1985-1998, have revealed that the $B$-value is decreased from the all-time value $B\simeq 0.88$ to $B\simeq 0.60$ before the mainshock [\textit{Kawamura}, 2006].  Such a change of the $B$-value observed in real seismicity turns out to be close even quantitatively to the one observed here in Fig.15(a).

 In Figs.16, we also show  the time-resolved local magnitude distributions {\it after\/} the mainshock for the cases of $\alpha =1$ (a), $\alpha=13$ (b) and $\alpha =20$ (c) with fixing $l=3$ and $\sigma=0.01$, each corresponding to the ``supercritical'', ``near-critical'' and ``subcritical'' regimes, respectively. Other conditions are the same as those in Figs.15. Similarly to Figs.15, an apparent $B$-value after the mainshock decreases or increases appreciably, depending on whether one is either in the ``supercritical'' or ``subcritical'' regime. Thus, the situation before and after the mainshock turn out to be quite similar. The analysis of the JUNEC catalogue yields a change of the $B$-value after the mainshock from $B\simeq 0.71$  to the all-time value $B\simeq 0.88$ [\textit{Kawamura}, 2006]. Again, this turns out to be rather close to our result of Fig.16(a) where the $B$-value has changed from $B\simeq 0.67$ to the all-time value $B\simeq 0.89$.

\section{Summary and discussion}

 Spatiotemporal correlations of the two-dimensional BK model of
 earthquakes were studied in a wide parameter range by means of
 extensive numerical computer simulations. A phase diagram is
 constructed in the plane of the elastic-parameter $l$ and the
 frictional-parameter $\alpha$ (Fig.4). The statistical properties of
 the model turn out to depend on the frictional-parameter $\alpha$
 most sensitively. The model exhibits a ``subcritical'' behavior for
 smaller and larger values of $\alpha$, while it exhibits a
 ``supercritical'' behavior for intermediate values of
 $\alpha$. Transition between the small-$\alpha$ ``subcritical''
 regime  and the ``supercritical'' regime is  discontinuous
 , while the one between  the ``supercritical''
 regime and the  large-$\alpha$ ``subcritical'' regime is continuous.

 At smaller magnitudes $\mu \lsim \tilde \mu$, the magnitude distribution $R(\mu)$ exhibits a  behavior close to the power law. At larger magnitudes  $\mu \gsim \tilde \mu$, however, $R(\mu)$ exhibits a significant deviation from the power law. In the ``supercritical'' regime realized at intermediate values of $\alpha$, $R(\mu)$ exhibits a pronounced peak structure at a large magnitude, falling off rapidly at still larger magnitudes, while, in the ``subcritical'' regimes realized at smaller and larger values of $\alpha$, $R(\mu)$ exhibits a decay faster than a power law without a peak structure. ``Near-critical'' behavior corresponding to the GR law is observed only near the boundary between the ``supercritical'' and the large-$\alpha$ ``subcritical'' regimes. In this sense, a true power-law distribution is not a generic attribute of the 2D BK model, neither in the 1D BK model. In other words, the size distribution of the BK model exhibit an off-critical or characteristic behavior at larger magnitudes, the GR-like ``near-critical'' behavior being observed only at smaller magnitudes. 

 Given its magnitude, the smaller-$\alpha$ ``subcritical'' behavior is characterized by smaller displacement, smaller stress-drop and larger rupture-zone size, while the larger-$\alpha$ ``subcritical'' behavior is characterized by larger displacement, larger stress-drop and smaller rupture-zone size.  Seismic events in the ``subcritical'' regime exhibit an interesting scaling property. Namely, the mean displacement at an event $\Delta \bar u$ scales with the seismic moment $M= \exp \mu$ as $\Delta \bar u\propto M^{0.1}$, while the mean number of failed-blocks $\bar N_b$ (the rupture-zone size) scales as $\bar N_b \propto M^{0.9}$. Seismic events in the ``supercritical'' regime exhibit a crossover behavior between the two ``subcritical'' behaviors at smaller magnitudes, but exhibit the scaling property at larger magnitudes.

 We also observed several intriguing precursory phenomena associated with the mainshock, particularly in the ``supercritical'' regime.  Preceding the mainshock, the frequency of smaller events is gradually enhanced around the epicenter of the upcoming mainshock. Such an increase of the seismic activity arises only in the ``supercritical'' regime, but is hardly visible in the ``subcritical'' regime. Following the seismic acceleration, the frequency of smaller events is dramatically suppressed just before the mainshock in a close vicinity of the epicenter of the upcoming mainshock, while it remains to be active in the surroundings (the Mogi doughnut). Such a doughnut-like quiescence is observed both in the ``subcritical'' and ``supercritical'' regimes in common, and can be accounted for as an instability induced by single-block events. On the other hand, the Omori law of aftershocks is not realized in the present 2D BK model, neither in the 1D BK model.

 An apparent $B$-value before the mainshock decreases or increases appreciably, depending on whether one is either in the ``supercritical'' or ``subcritical'' regime. Almost the same change of an apparent $B$-value is observed also after the mainshock. Our result might give a hint in understanding the observed change of the $B$-value in real earthquakes. 

 The existence of these distinct precursory phenomena in the 2D BK model may open a way to the prediction of the time and the position of the upcoming large event.  Of course, the BK model is a highly simplified model of real earthquake faults, {\it e.g.\/}, finite discretization of the originally continuum crust being made, highly simplified constitutive relation being assumed, {\it etc\/}. Furthermore, although the present 2D model is certainly more realistic than the corresponding 1D model in that the full 2D fault plane has been taken into account, the real earthquake fault might not necessarily be a sharp 2D plane, but rather be a more fractal object with complex branch structure: See, {\it e.g.\/}, [\textit{Kagan}, 2006]. Nevertheless, we hope that some of the fundamental characteristics of seismic events of the 2D BK model as revealed here might provide us with a useful reference in understanding the properties of real seismicity.

\setfigurenum{1}
\begin{figure}[ht]
\begin{center}
\includegraphics[scale=0.65]{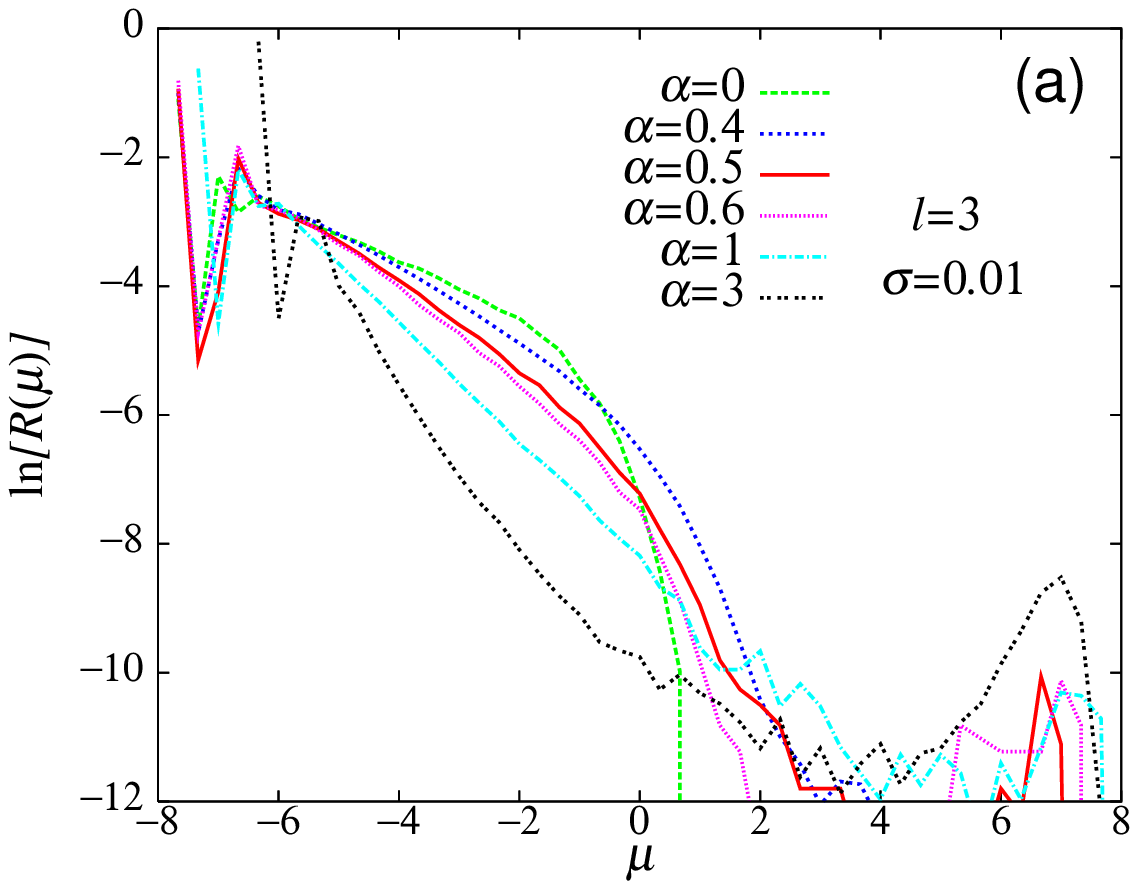}
\includegraphics[scale=0.65]{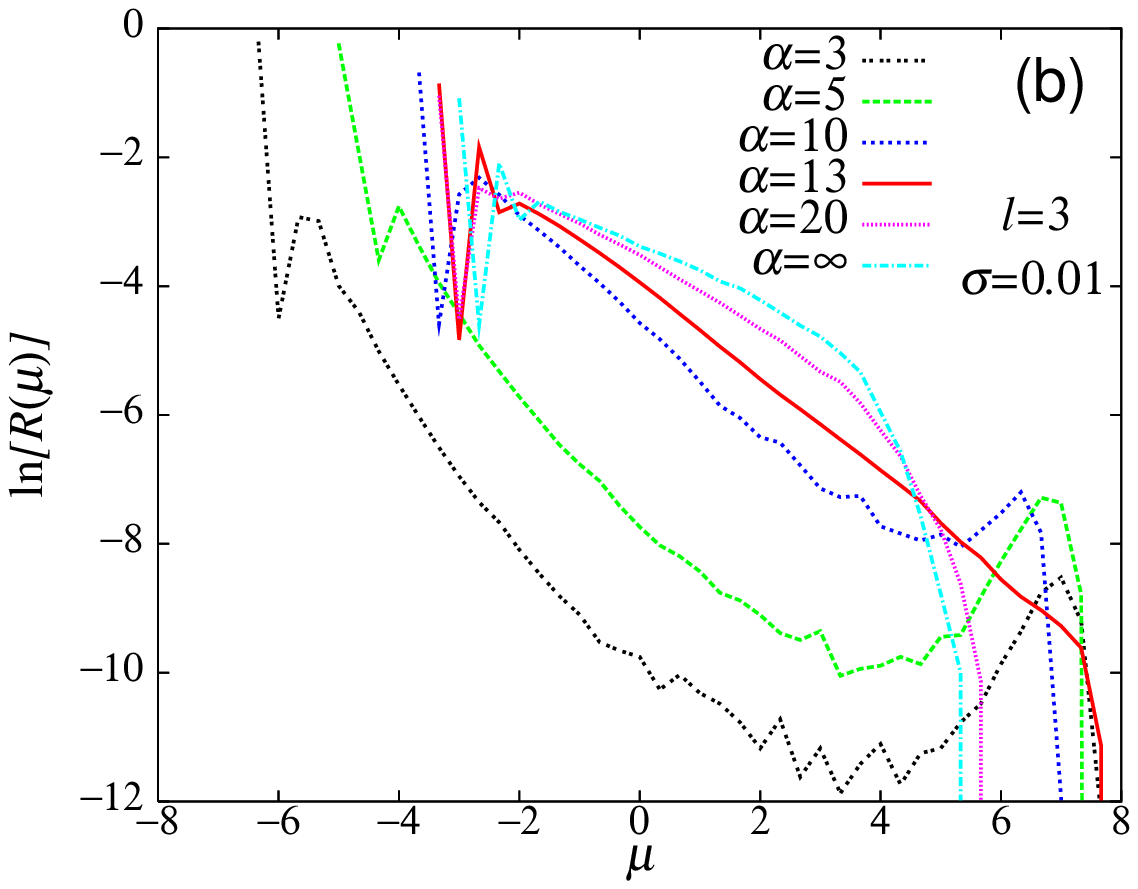}
\includegraphics[scale=0.65]{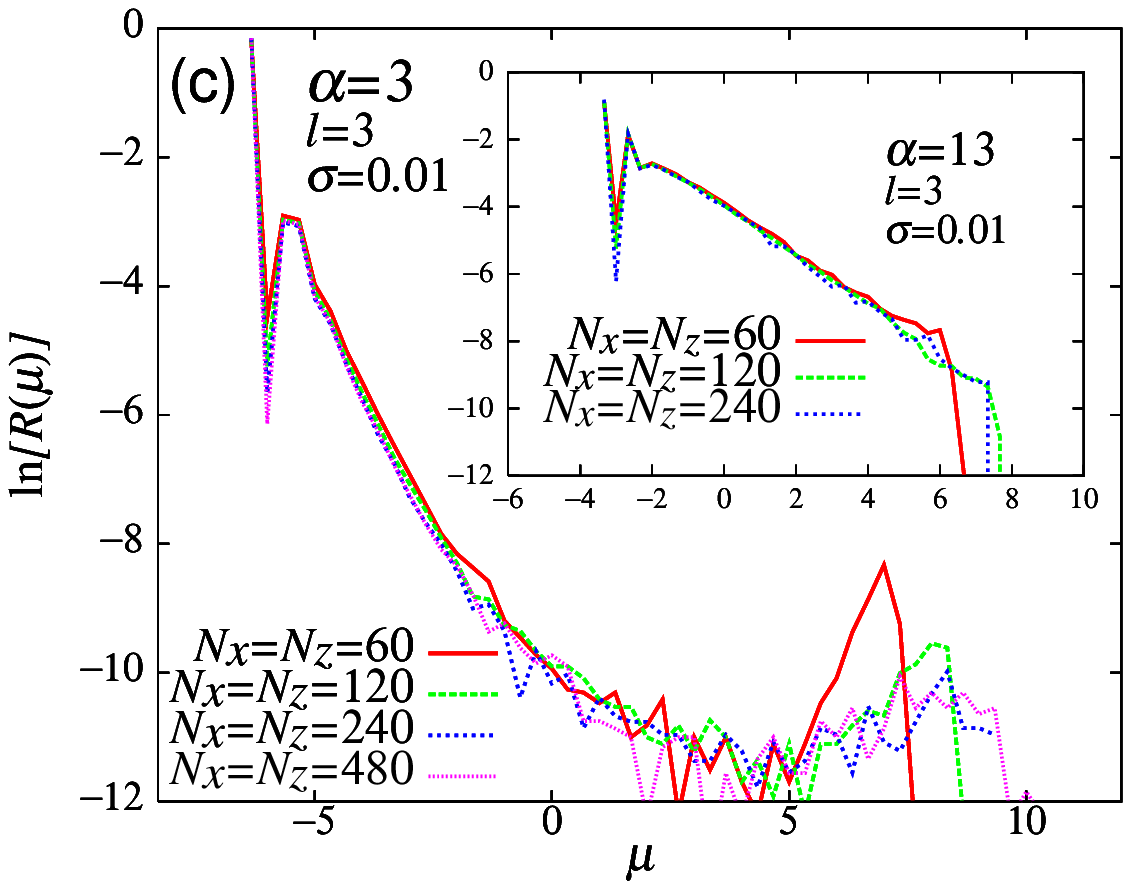}
\end{center}
\caption{
Magnitude distribution $R(\mu)$ of earthquake events for the  parameters $l=3$ and $\sigma=0.01$. Fig.(a) represents $R(\mu)$ for  smaller values of the frictional-parameter $0\leq \alpha \leq 3$, while Fig.(b) represents $R(\mu)$ for larger values of the frictional-parameter $3\leq \alpha \leq \infty$. The system size is $60\times 60$. In Fig.(c), the size dependence of $R(\mu)$ is shown for the case of $\alpha=3$ (main panel) and $\alpha=13$ (inset).
}
\end{figure}

\setfigurenum{2}
\begin{figure}[ht]
\begin{center}
\includegraphics[scale=0.65]{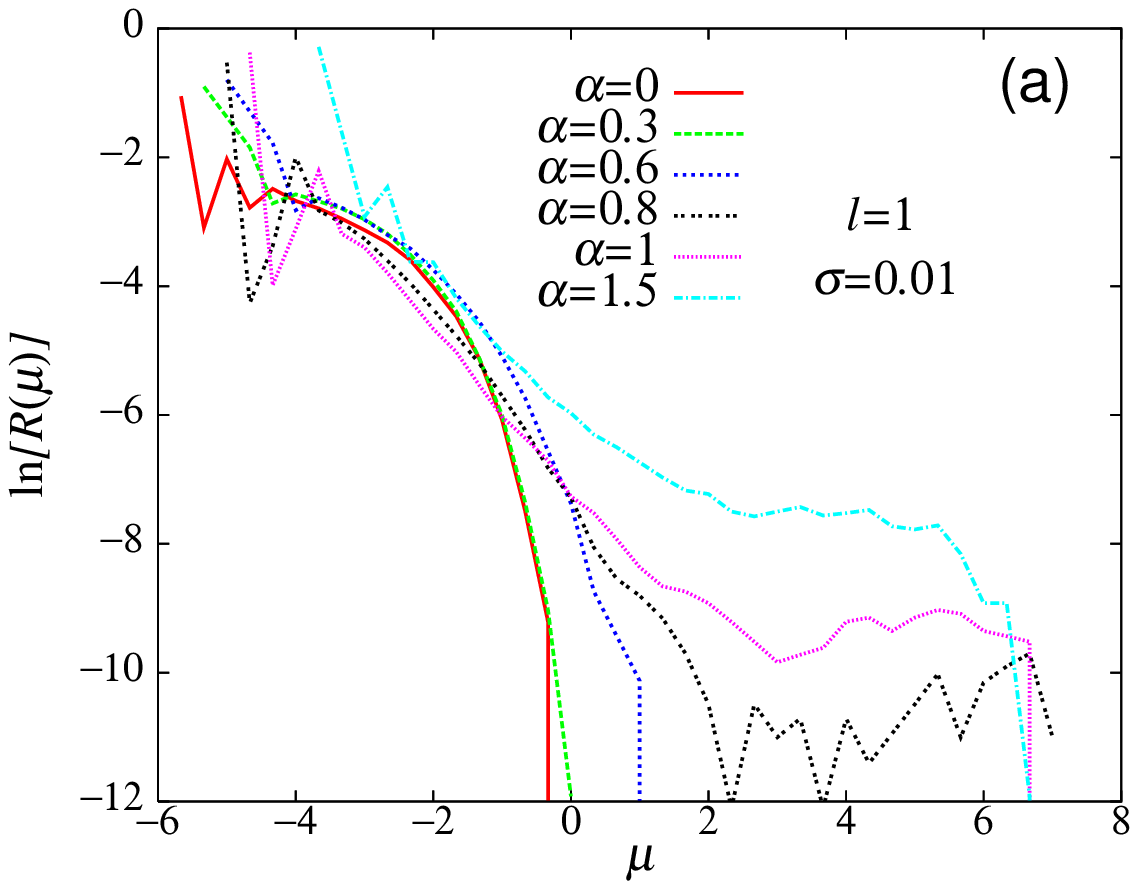}
\includegraphics[scale=0.65]{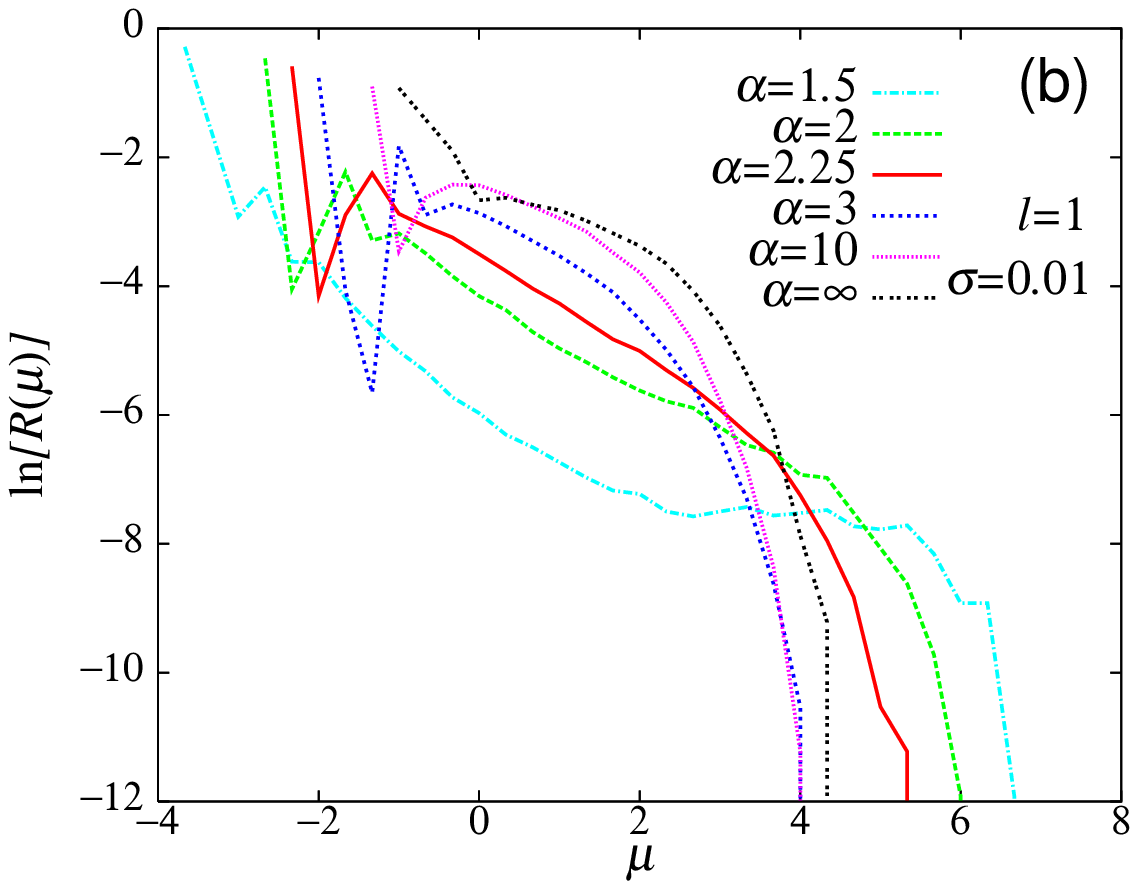}
\end{center}
\caption{
Magnitude distribution $R(\mu)$ of earthquake events for the parameters $l=1$ and $\sigma=0.01$. Fig.(a) represents $R(\mu)$ for  smaller values of the frictional-parameter $0\leq \alpha \leq 1.5$, while Fig.(b) represents $R(\mu)$ for larger values of the frictional-parameter $1.5\leq \alpha \leq \infty$. The system size is $60\times 60$. 
}
\end{figure}

\setfigurenum{3}
\begin{figure}[ht]
\begin{center}
\includegraphics[scale=0.65]{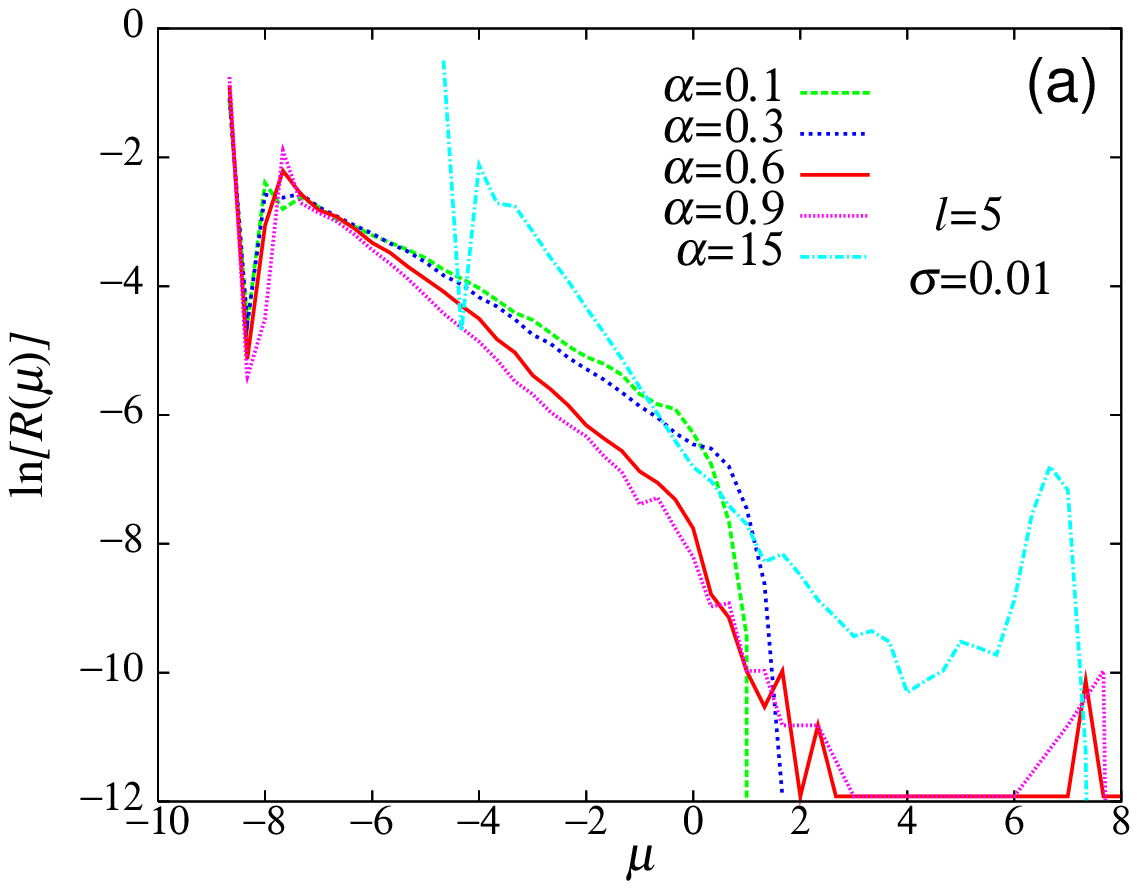}
\includegraphics[scale=0.65]{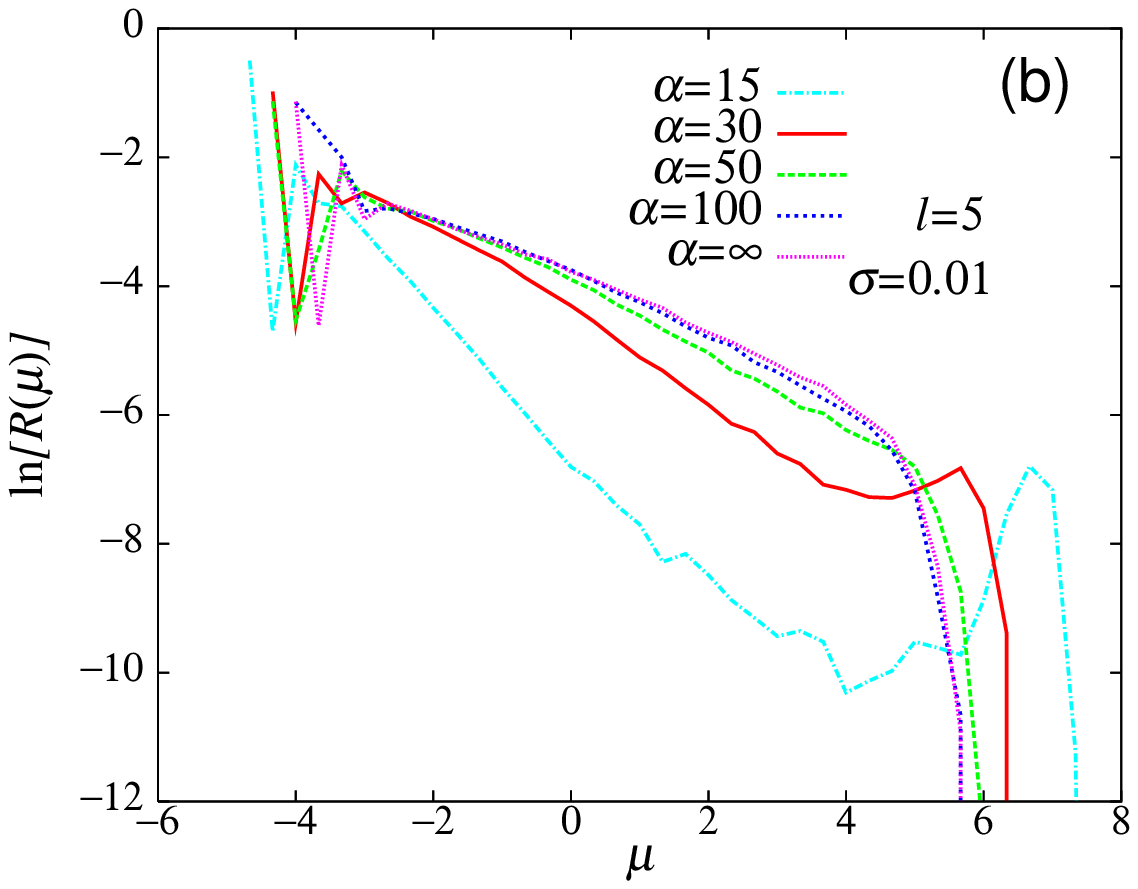}
\end{center}
\caption{
Magnitude distribution $R(\mu)$ of earthquake events for the parameters $l=5$ and $\sigma=0.01$. Fig.(a) represents $R(\mu)$ for  smaller values of the frictional-parameter $0\leq \alpha \leq 15$, while Fig.(b) represents $R(\mu)$ for larger values of the frictional-parameter $15\leq \alpha \leq \infty$. The system size is $60\times 60$.
}
\end{figure}

\setfigurenum{4}
\begin{figure}[ht]
\begin{center}
\includegraphics[scale=0.65]{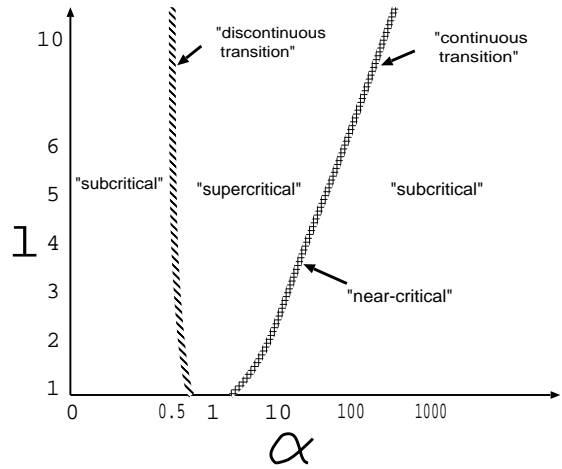}
\end{center}
\caption{
Phase diagram in the frictional-parameter $\alpha$ versus the elastic-parameter $l$ plane. The parameter $\sigma$ is  $\sigma=0.01$.
}
\end{figure}

\setfigurenum{5}
\begin{figure}[ht]
\begin{center}
\includegraphics[scale=0.65]{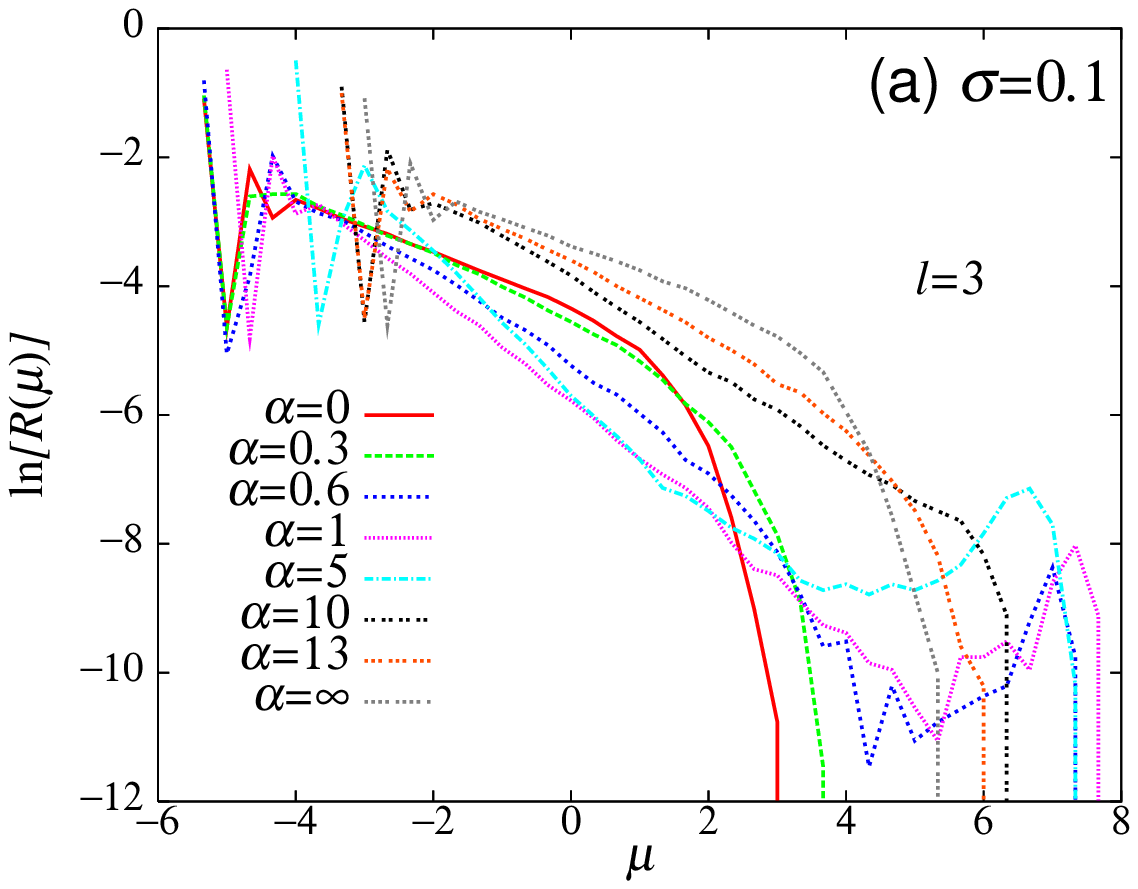}
\includegraphics[scale=0.65]{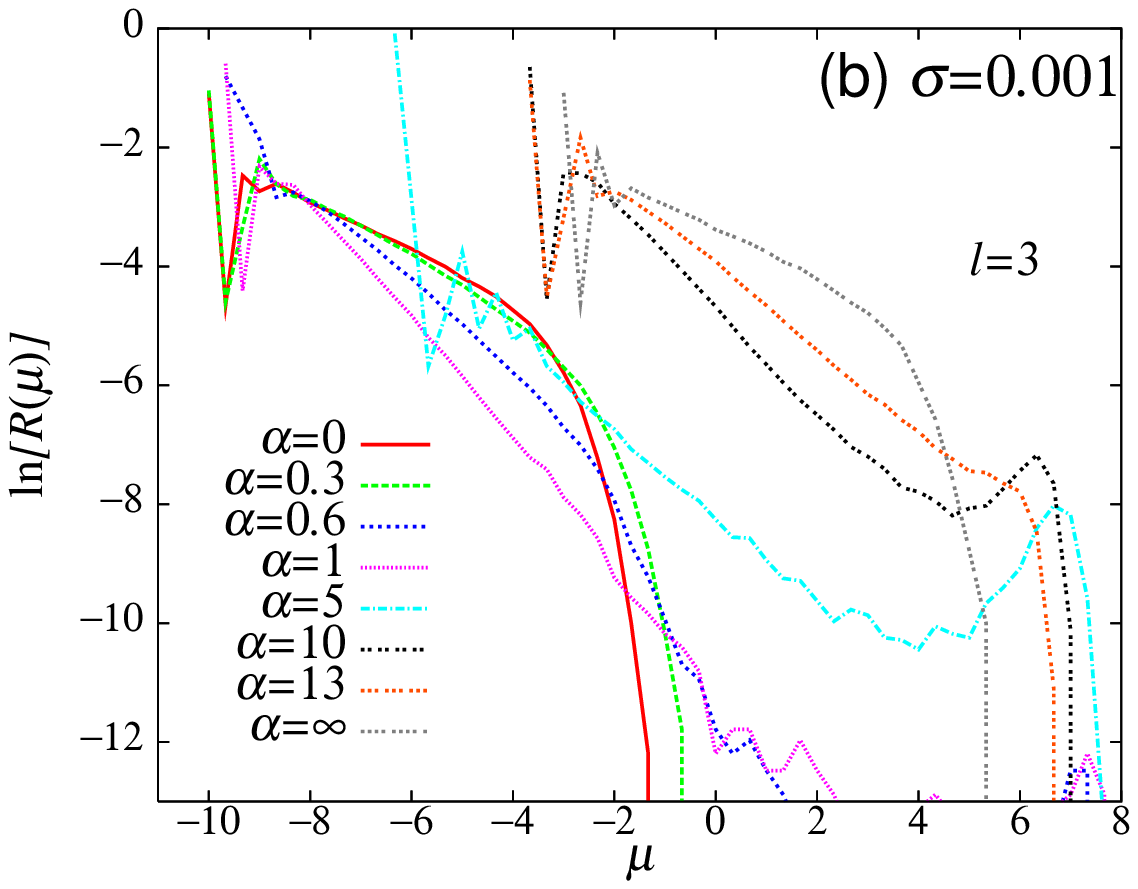}
\end{center}
\caption{
Magnitude distribution $R(\mu)$ of earthquake events for various values of $\alpha$ in the case of the frictional-parameter $\sigma=0.1$ (a), and $\sigma=0.001$ (b). The elastic parameter is $l=3$.  The system size is $60\times 60$. 
}
\end{figure}

\setfigurenum{6}
\begin{figure}[ht]
\begin{center}
\includegraphics[scale=0.65]{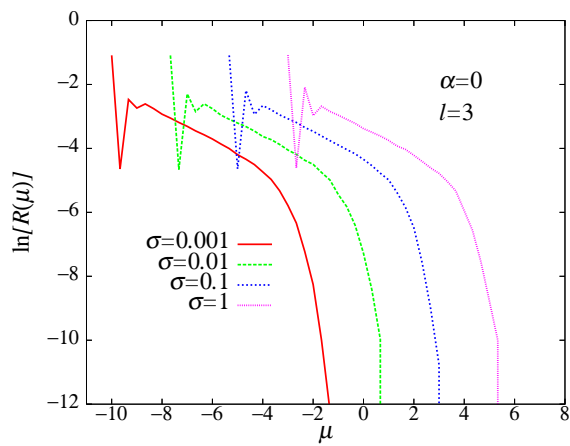}
\end{center}
\caption{
Magnitude distribution $R(\mu)$ of earthquake events for various values of the frictional-parameter $\sigma$. The parameter $\alpha$ is fixed to $\alpha=0$, which corresponds to a constant dynamical friction of strength $1-\sigma$. The system size is $60\times 60$.
}
\end{figure}

\setfigurenum{7}
\begin{figure}[ht]
\begin{center}
\includegraphics[scale=0.65]{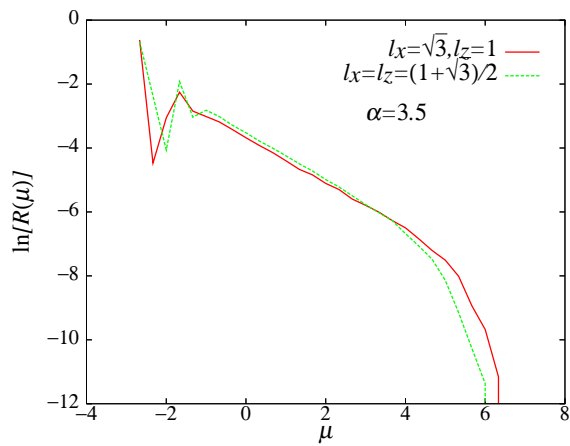}
\end{center}
\caption{
Magnitude distribution $R(\mu)$ of earthquake events in the case of spatially anisotropic elastic parameters, $l_x=\sqrt 3$ and $l_z=1$,  compared with the one in the case of spatially isotropic elastic parameters with the same mean value, {\it i.e.\/}, $l_x=l_z=(1+\sqrt 3)/2$. The parameters $\alpha$ and $\sigma$ are $\alpha=3.5$ and $\sigma=0.01$. The system size is $160\times 80$. 
}
\end{figure}

\setfigurenum{8}
\begin{figure}[ht]
\begin{center}
\includegraphics[scale=0.65]{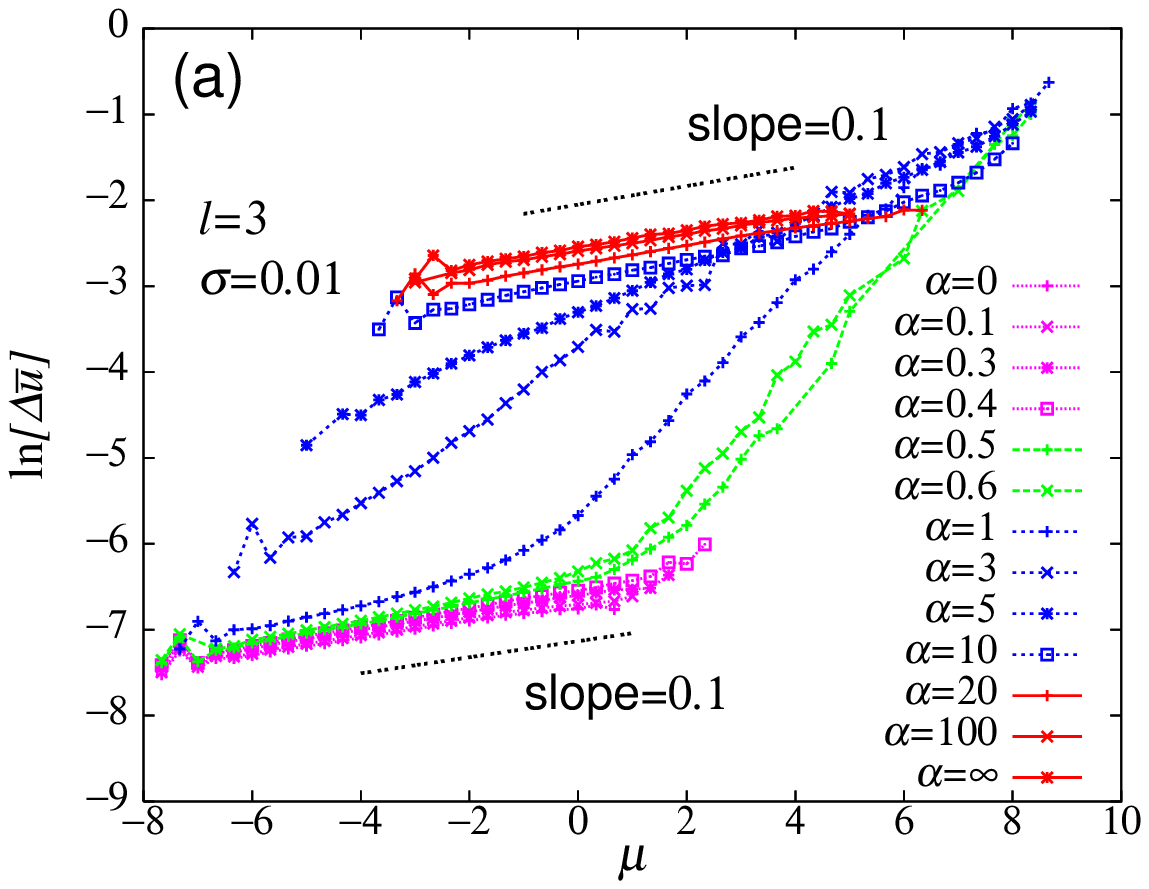}
\includegraphics[scale=0.65]{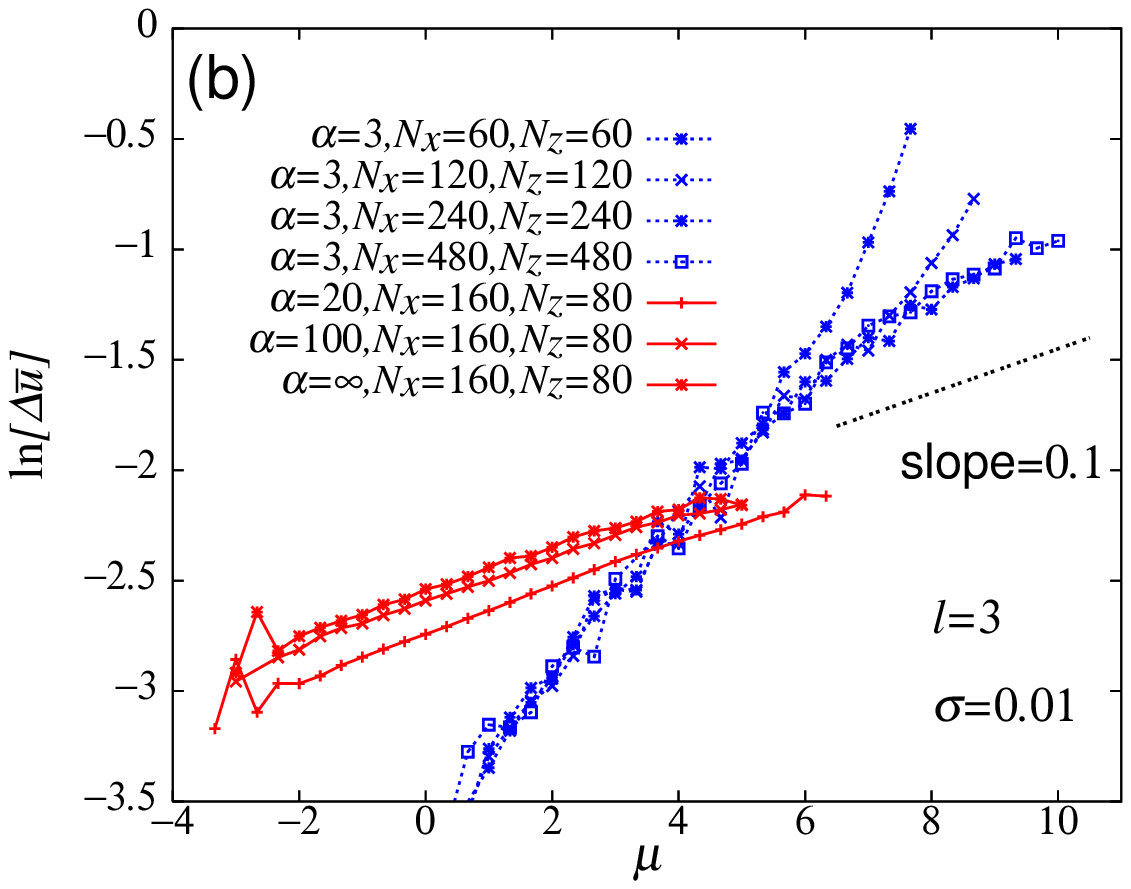}
\end{center}
\caption{
(a) The magnitude dependence of the mean displacement  for various values of the frictional-parameter $\alpha$. The parameters $l$ and $\sigma$ are fixed to $l=3$ and $\sigma=0.01$. Small $\alpha$ ($\alpha\lsim 0.5$) and large $\alpha$ ($\alpha\gsim 13$) correspond to the ``subcritical'' regimes, while intermediate $\alpha$ ($0.5\lsim \alpha\lsim 13$) corresponds to the ``supercritical'' regime. The data in the different regimes, {\it i.e.\/}, the small-$\alpha$ ``subcritical'' regime, the ``discontinuous transition'' regime,  the ``supercritical'' regime and the large-$\alpha$ ``subcritical'' regimes, are indicated by different colors in the figures. The system size is $160\times 80$. (b) The mean displacement at larger magnitudes are shown on an expanded scale for the $N\times N$ systems, with varying $N$ in the range $60\leq N\leq 480$.
}
\end{figure}

\setfigurenum{9}
\begin{figure}[ht]
\begin{center}
\includegraphics[scale=0.65]{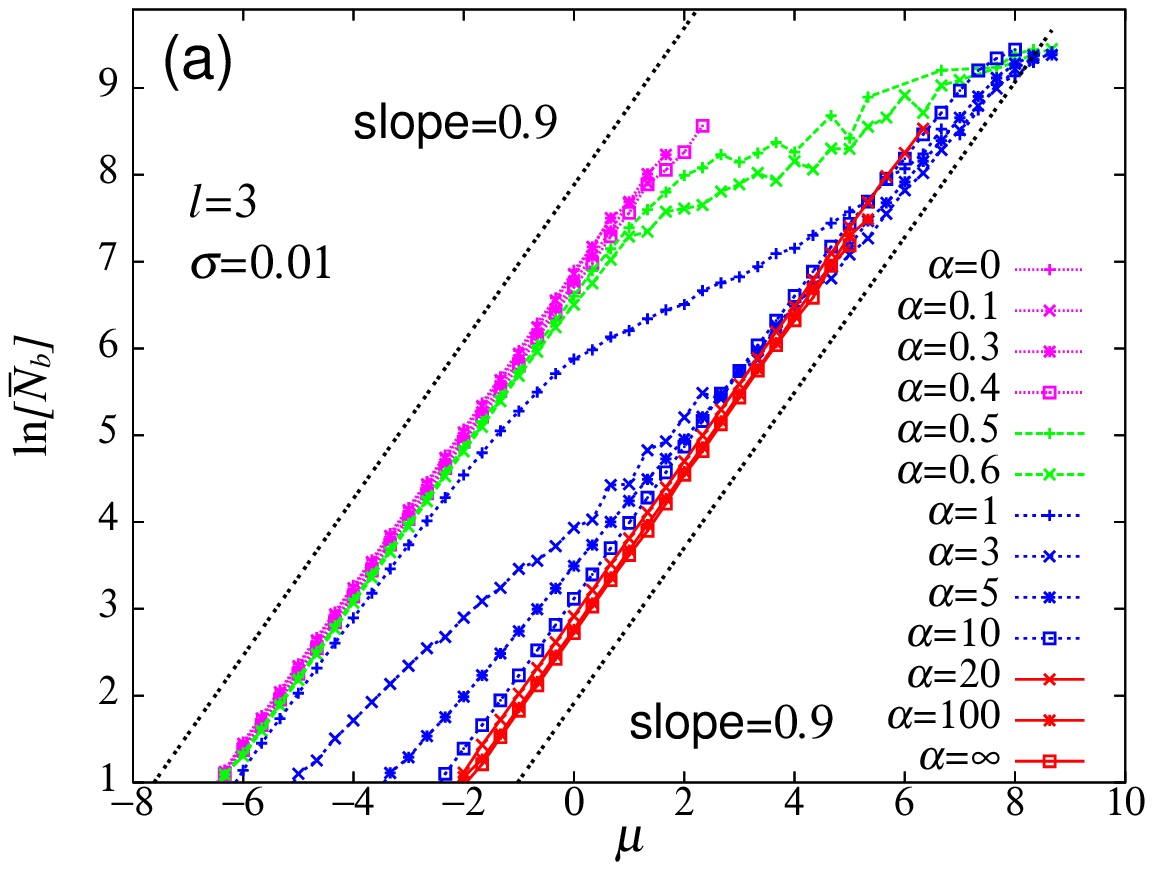}
\includegraphics[scale=0.65]{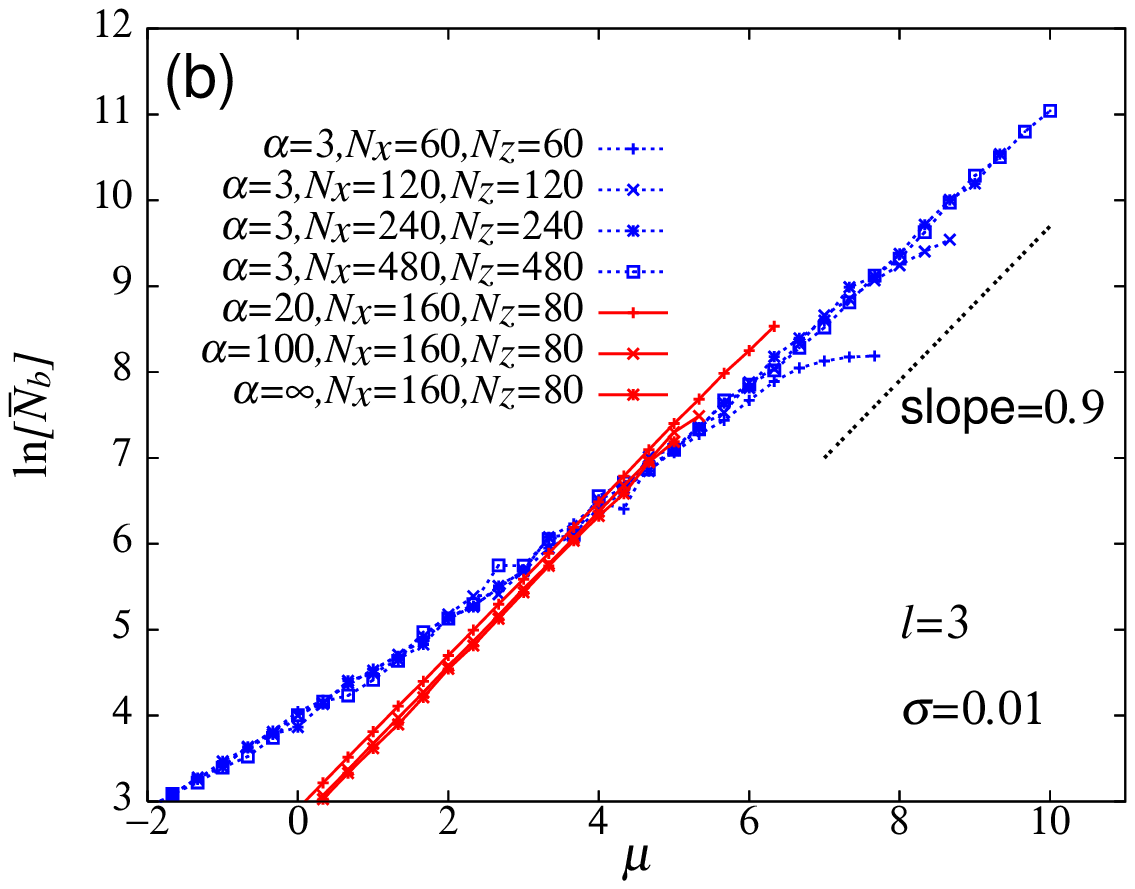}
\end{center}
\caption{
(a) The magnitude dependence of  the mean number of failed-blocks of each event for various values of the frictional-parameter $\alpha$. The parameters $l$ and $\sigma$ are fixed to $l=3$ and $\sigma=0.01$. Small $\alpha$ ($\alpha\lsim 0.5$) and large $\alpha$ ($\alpha\gsim 13$) correspond to the ``subcritical'' regimes, while intermediate $\alpha$ ($0.5\lsim \alpha\lsim 13$) corresponds to the ``supercritical'' regime. The data in the different regimes, {\it i.e.\/}, the small-$\alpha$ ``subcritical'' regime, the ``discontinuous transition'' regime,  the ``supercritical'' regime and the large-$\alpha$ ``subcritical'' regimes, are indicated by different colors in the figures. The system size is $160\times 80$. (b) The mean number of failed-blocks at larger magnitudes are shown on an expanded scale for the $N\times N$ systems, with varying $N$ in the range $60\leq N\leq 480$.
}
\end{figure}

\setfigurenum{10}
\begin{figure}[ht]
\begin{center}
\includegraphics[scale=0.65]{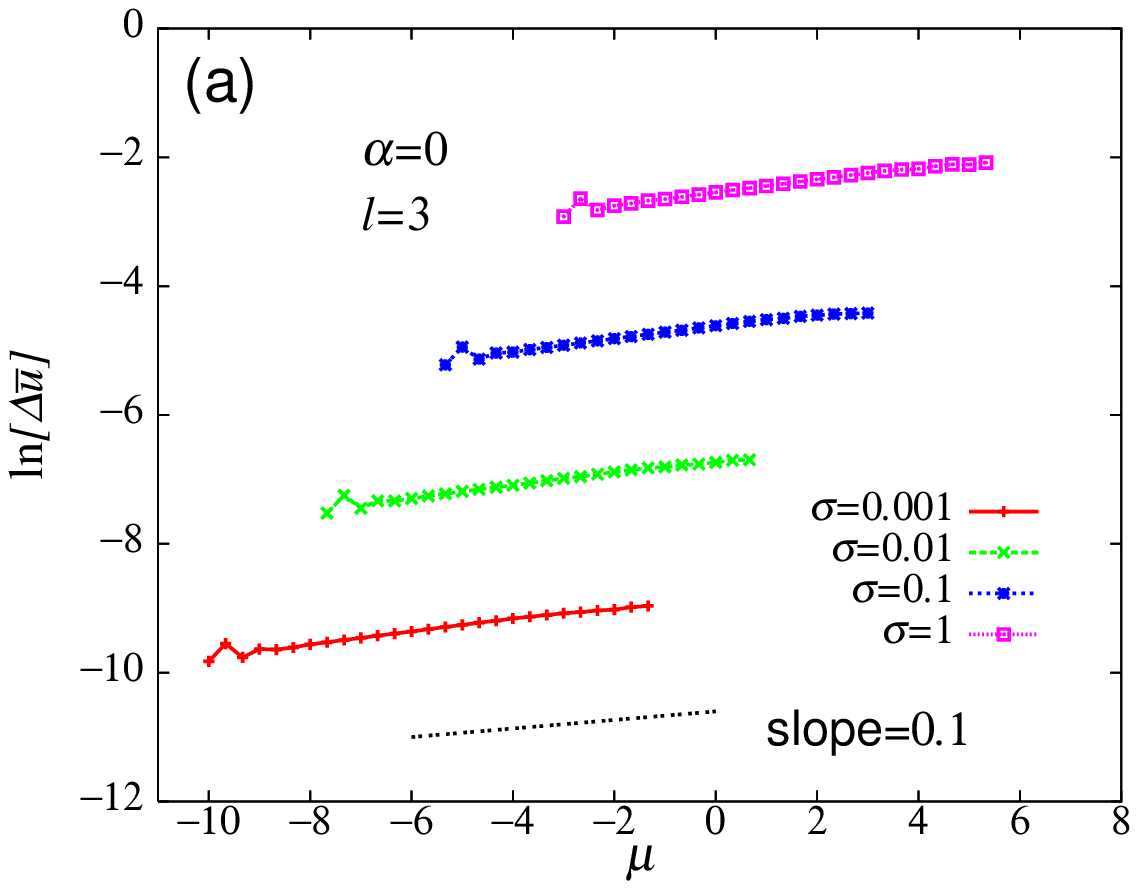}
\includegraphics[scale=0.65]{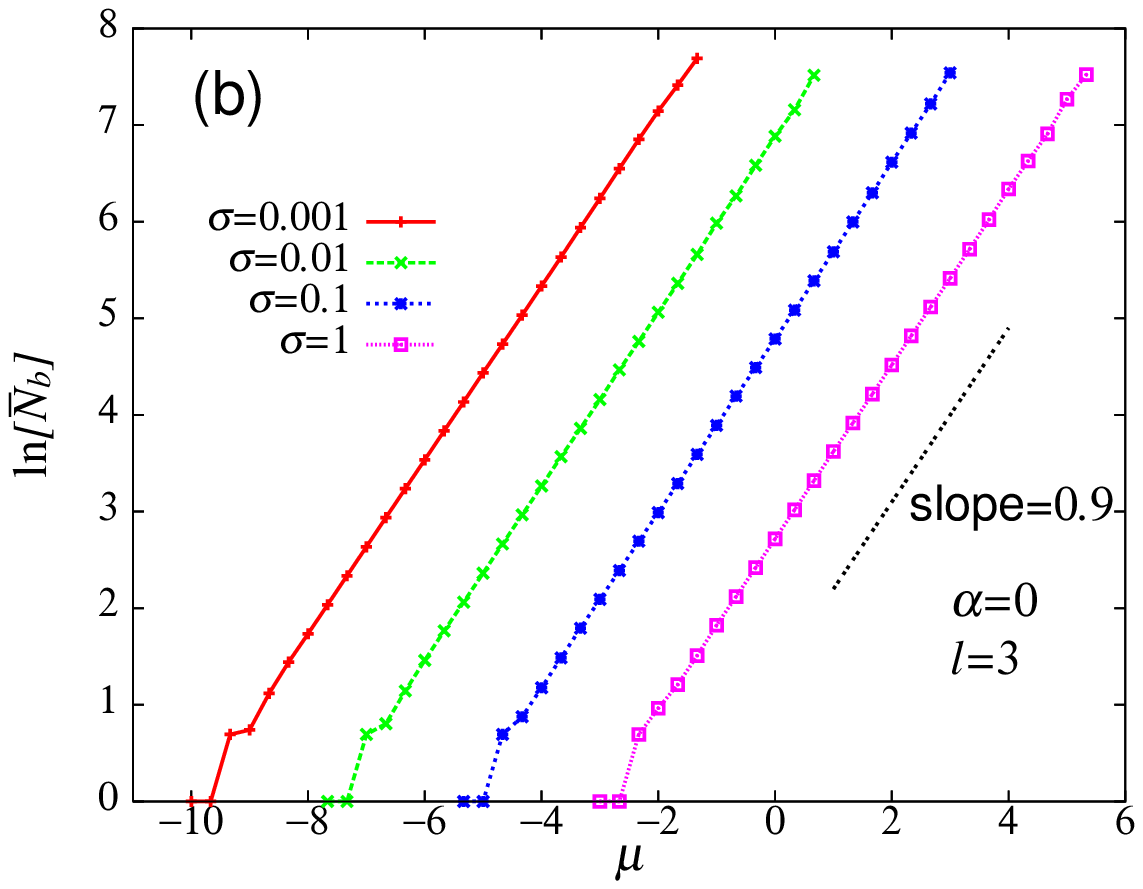}
\end{center}
\caption{
The magnitude dependence of the mean displacement (a), and of the mean number of failed-blocks (b) of each event, for various values of the frictional-parameter $\sigma$ in the case of $\alpha=0$, which corresponds to a constant dynamical friction of strength $1-\sigma$. The parameters $l$ is fixed to $l=3$. The system size is $160 \times 80$.
}
\end{figure}

\setfigurenum{11}
\begin{figure}[ht]
\begin{center}
\includegraphics[scale=0.65]{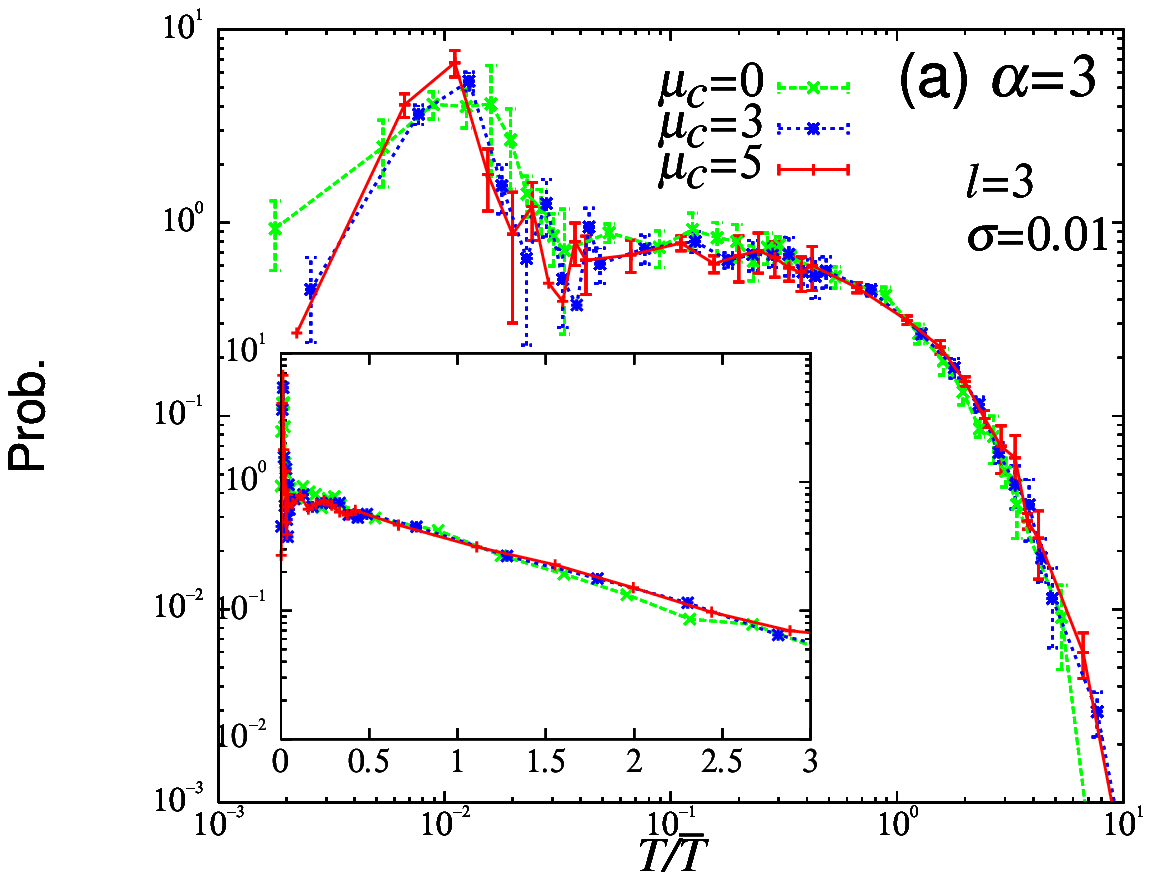}
\includegraphics[scale=0.65]{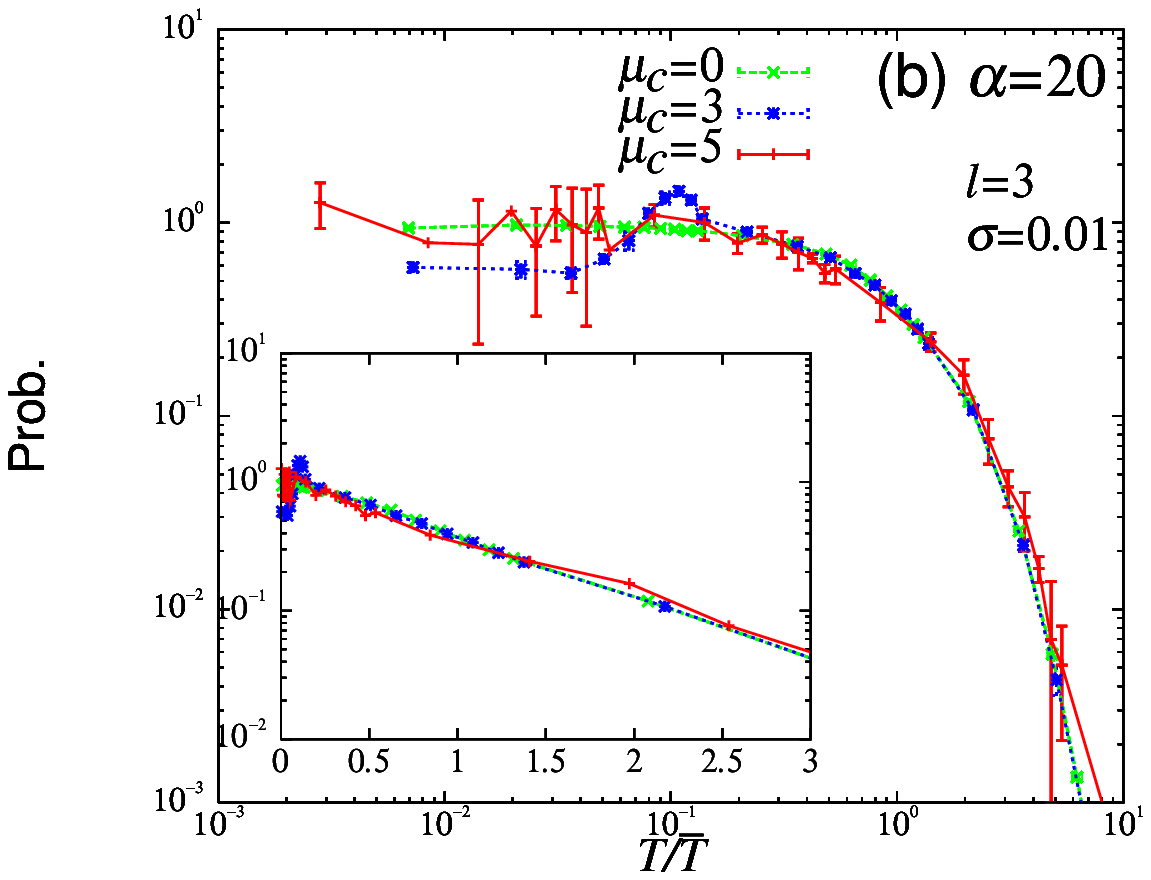}
\end{center}
\caption{
The local recurrence-time distribution of large events with their magnitudes
greater than $\mu_c$ ($\mu_c=5,3,0$) for $\alpha=3$ (a), and for $\alpha=20$ (b), each corresponding to the ``supercritical'' and ``subcritical'' regimes. The parameters $l$ and $\sigma$ are fixed  to be $l=3$ and $\sigma =0.01$. The system size is $160\times 80$. The recurrence time $T$ is normalized by its mean $\bar T$, which is $\bar T=14.0$, 19.5, 22.5, respectively for $\mu_c=0$, 3, 5 for the case of Fig.(a), and  $\bar T=0.14$, 1.38, 35.5, respectively for $\mu_c=0$, 3, 5 in the case of Fig.(b).
The insets represent the semi-logarithmic plots including the tail part of the distribution. The tail part shows an exponential behavior for both cases of $\alpha=3$ and 20.
}
\end{figure}

\setfigurenum{12}
\begin{figure}[ht]
\begin{center}
\includegraphics[scale=0.65]{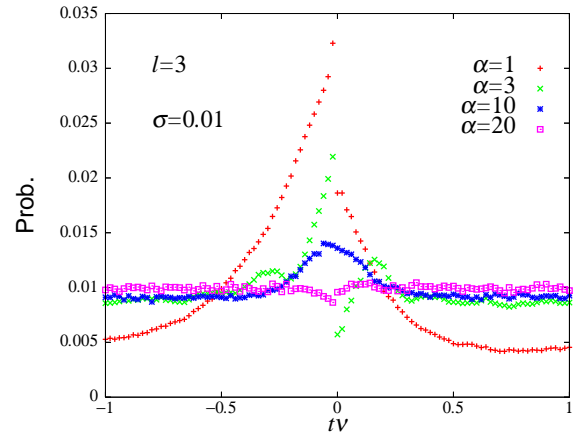}
\end{center}
\caption{
The time correlation function 
between large events with $\mu_c=5$ (mainshock) occurring at time $t=0$ and
events of arbitrary sizes (dominated in number by small events) occurring at
time $t$
for various values of $\alpha$. 
The parameters $l$ and $\sigma$ are fixed
 to be $l=3$ and $\sigma =0.01$.
Events of arbitrary sizes occurring within 5 blocks from the epicenter of the
mainshock are counted. 
The negative time $t<0$ represents the time before the mainshock, 
while the positive time $t>0$ represents the time after the mainshock. 
The average is taken over all large events with its magnitude $\mu >\mu_c=5$. The system size is $160\times 80$. 
}
\end{figure}

\setfigurenum{13}
\begin{figure}[ht]
\begin{center}
\includegraphics[scale=0.65]{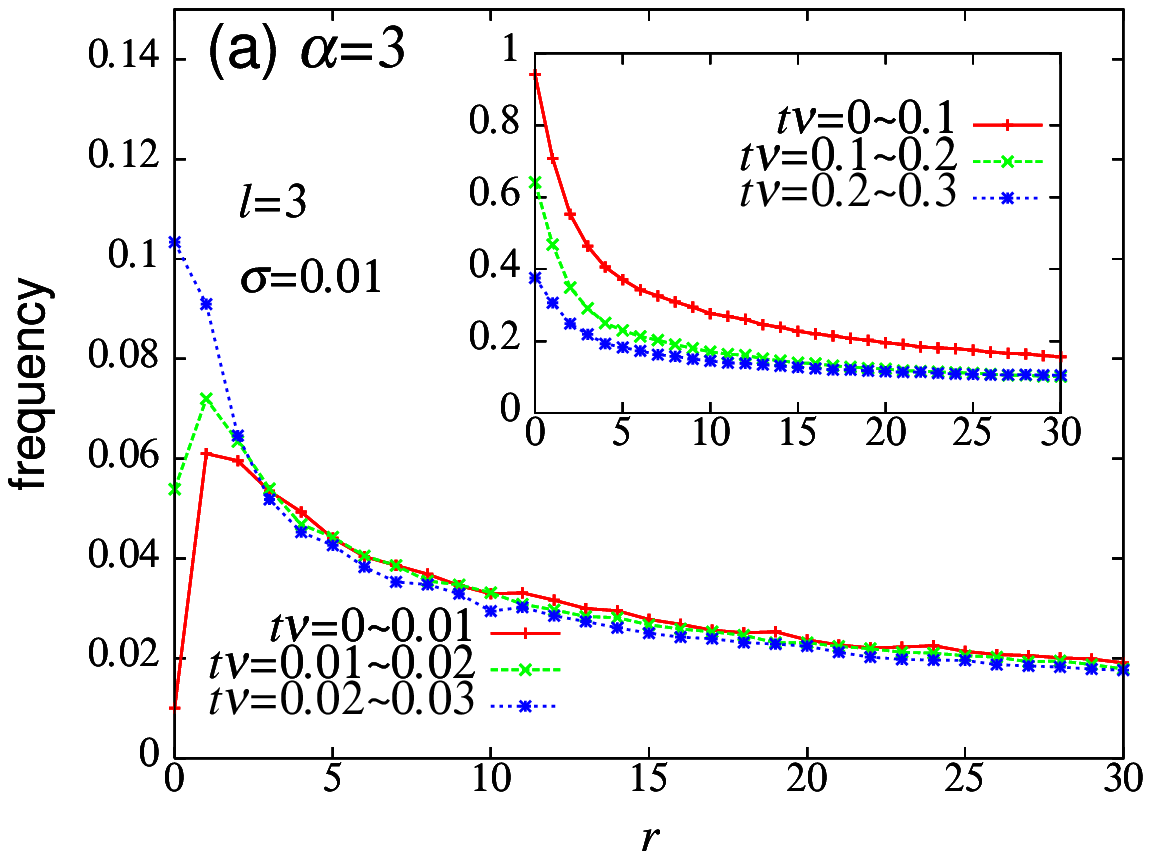}
\includegraphics[scale=0.65]{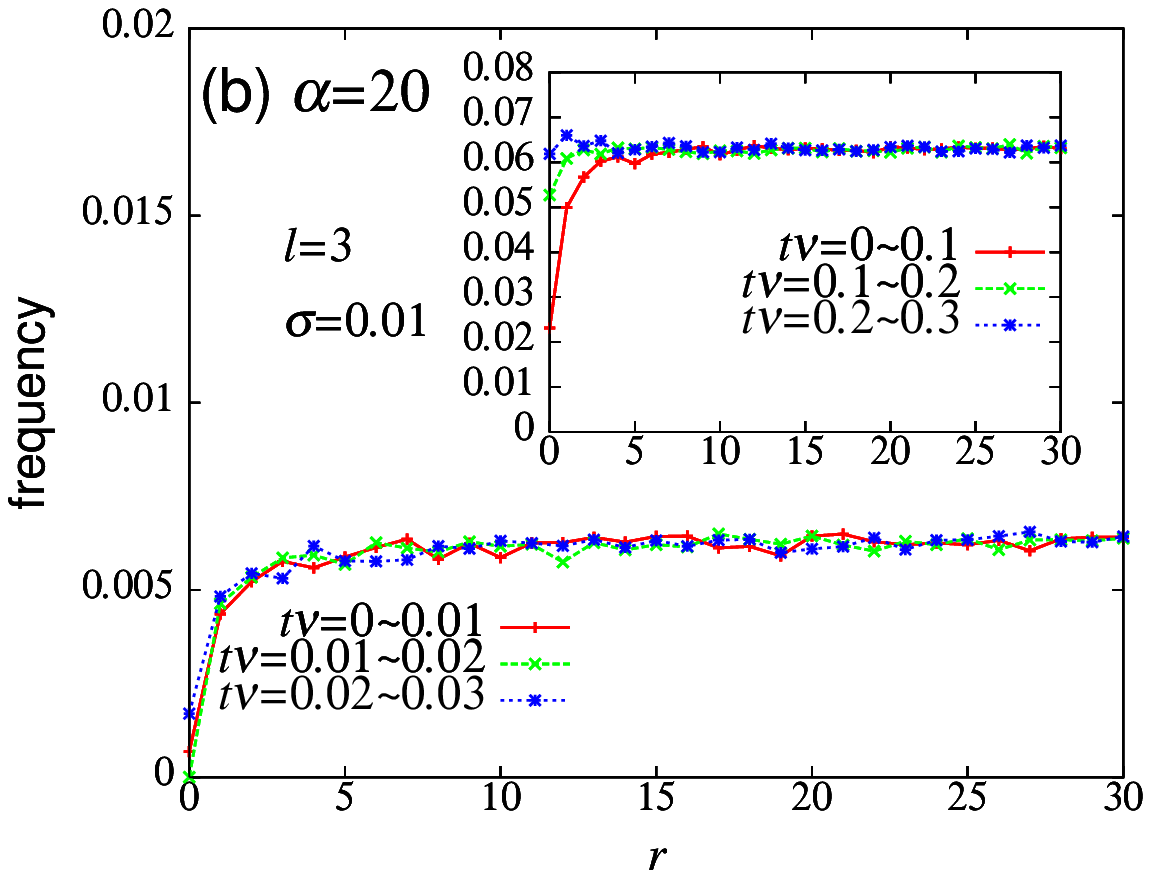}
\end{center}
\caption{
Event frequency preceding the large event with $\mu >\mu_c=5$  (mainshock) plotted versus $r$,
the distance from the epicenter of the upcoming mainshock for several time periods before the mainshock. The parameter $\alpha$ is 
$\alpha=3$ (a), and $\alpha=20$ (b) with $l=3$ and $\sigma=0.01$, each corresponding to the ``supercritical'' and ``subcritical'' regimes. The system size is $160\times 80$. The insets represent similar plots with longer-time intervals.
}
\end{figure}

\setfigurenum{14}
\begin{figure}[ht]
\begin{center}
\includegraphics[scale=0.65]{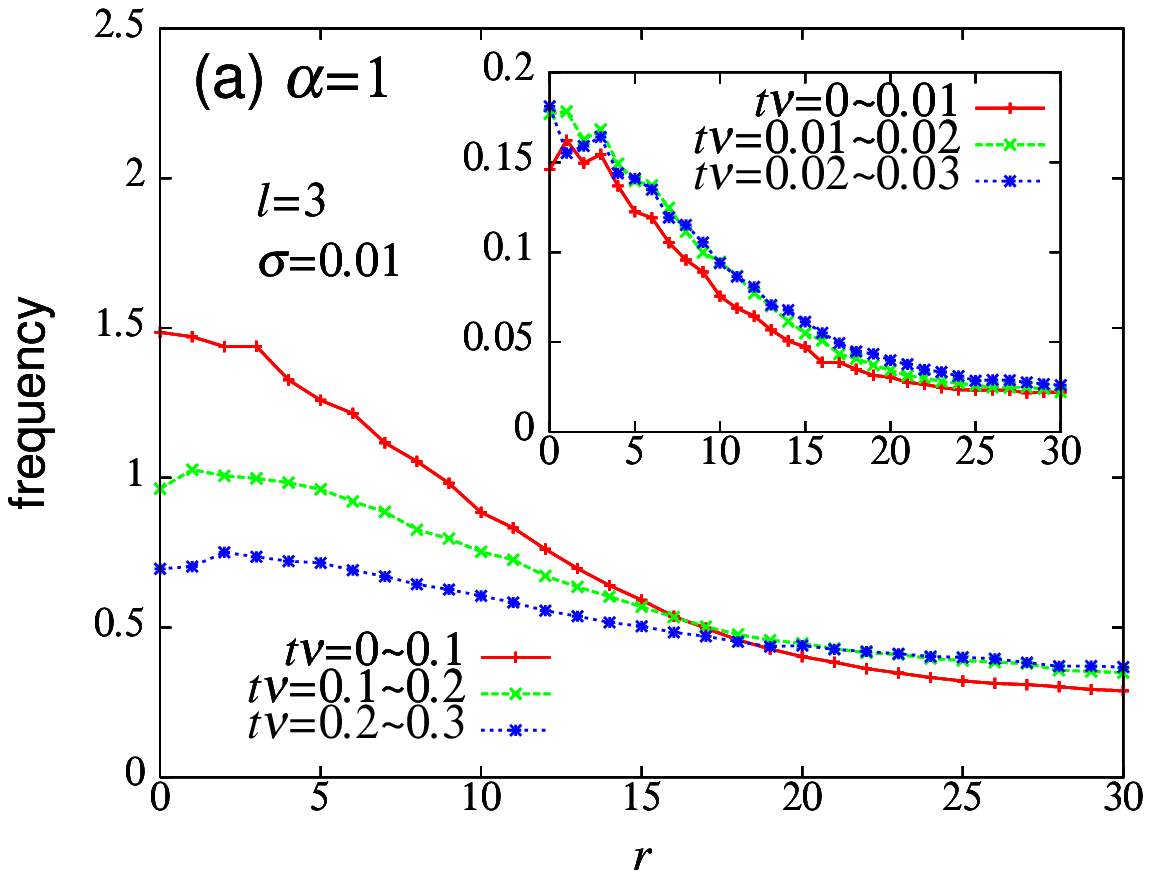}
\includegraphics[scale=0.65]{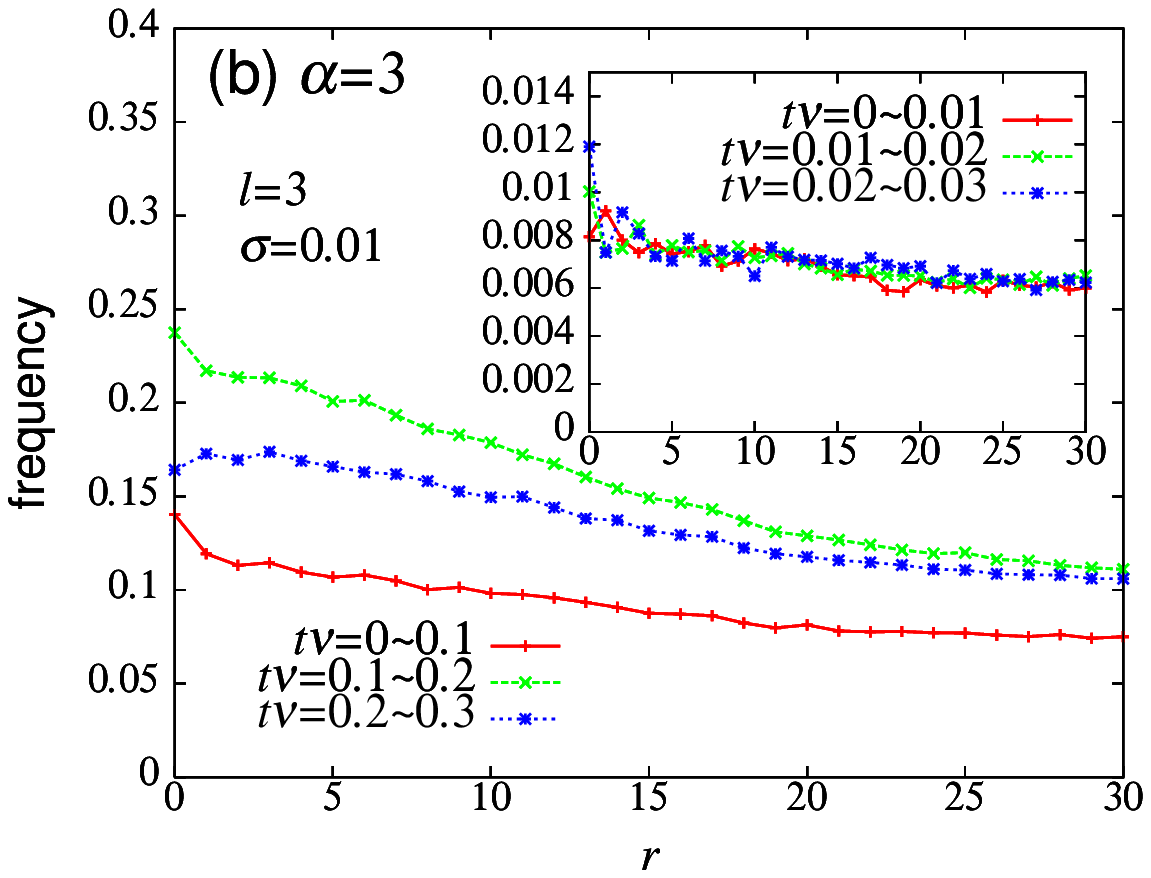}
\includegraphics[scale=0.65]{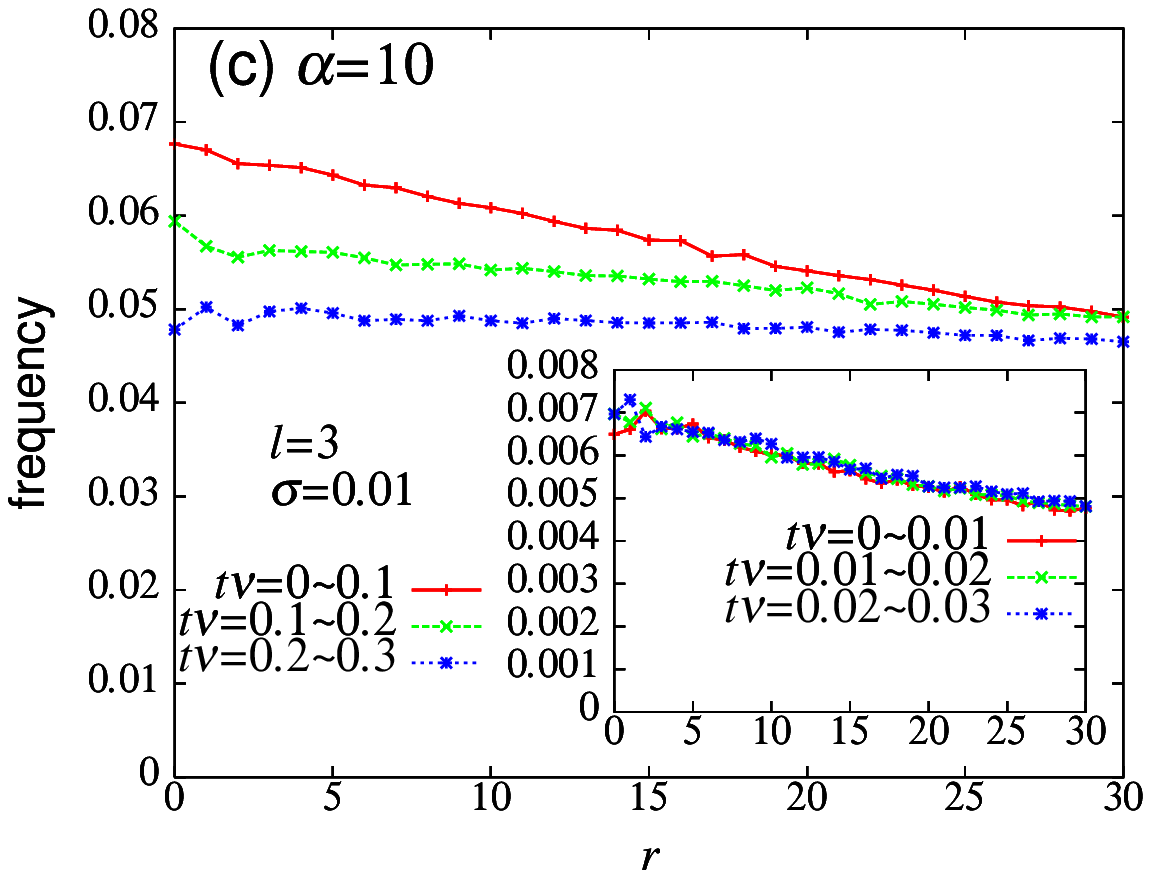}
\includegraphics[scale=0.65]{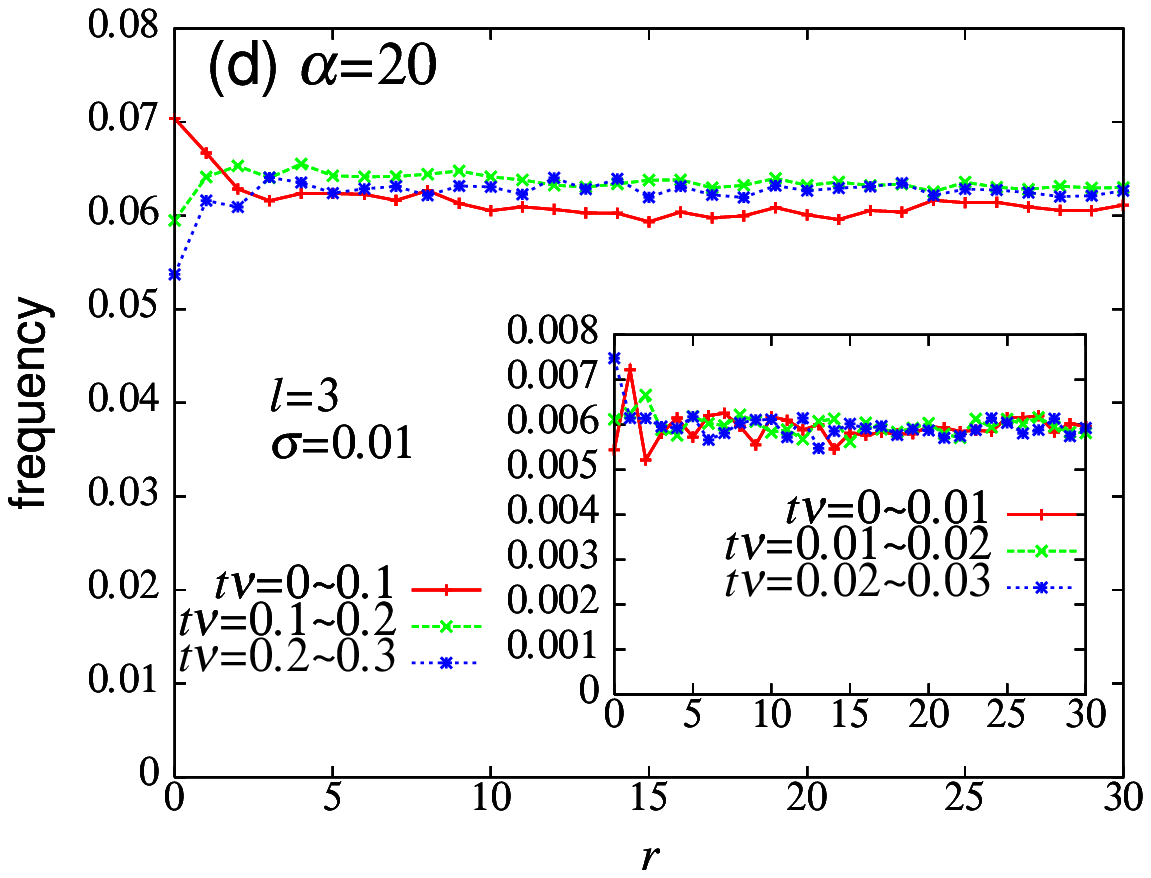}
\end{center}
\caption{
Event frequency after the large event with $\mu >\mu_c=5$  (mainshock) plotted
versus $r$,
the distance from the epicenter of the preceding mainshock, 
for several time periods after the mainshock. The parameter is 
$\alpha=1$ (a), $\alpha=3$ (b), $\alpha=10$ (c), and  $\alpha=20$ (d).
Figs.(a)-(c) correspond to the ``supercritical'' regime, while Fig.(d) to the ``subcritical'' regime. The parameters $l$ and $\sigma$ are fixed
 to be $l=3$ and $\sigma =0.01$. The system size is $160\times 80$. 
The insets  represent the shorter-time scale.
}
\end{figure}

\setfigurenum{15}
\begin{figure}[ht]
\begin{center}
\includegraphics[scale=0.65]{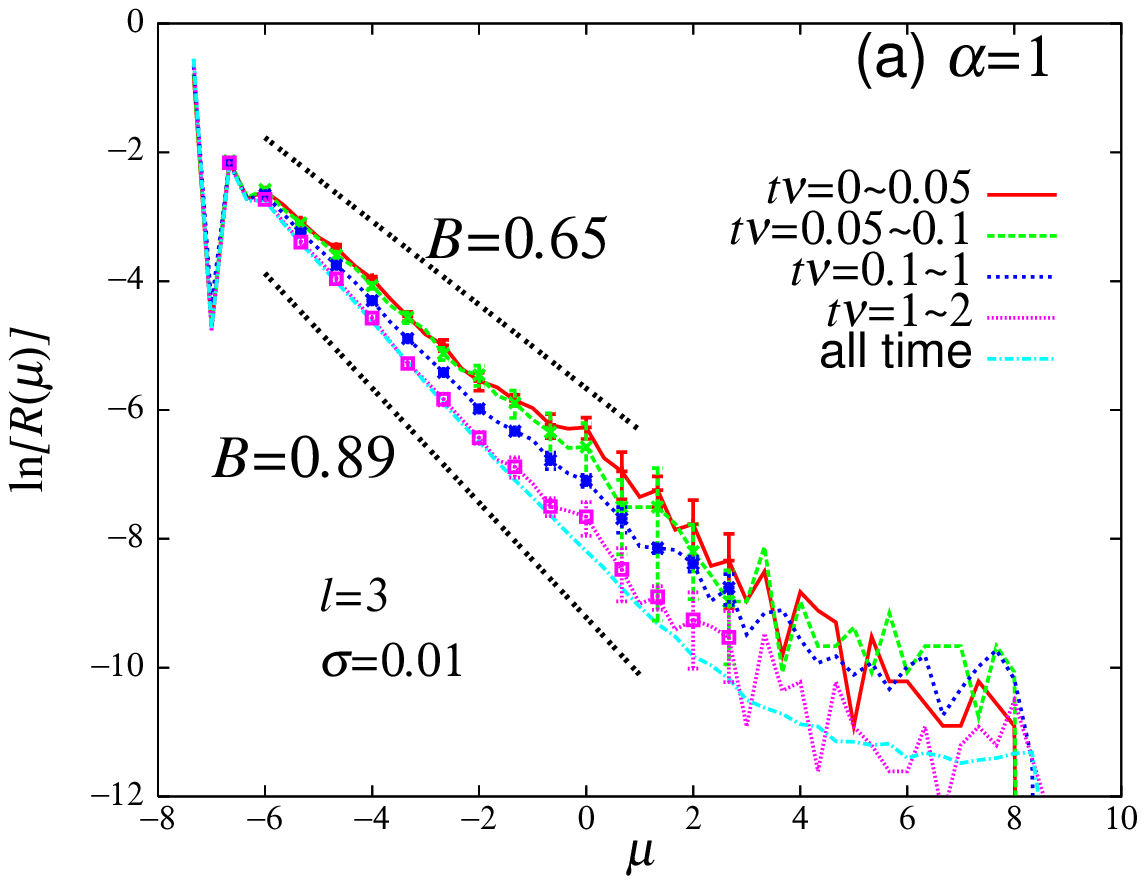}
\includegraphics[scale=0.65]{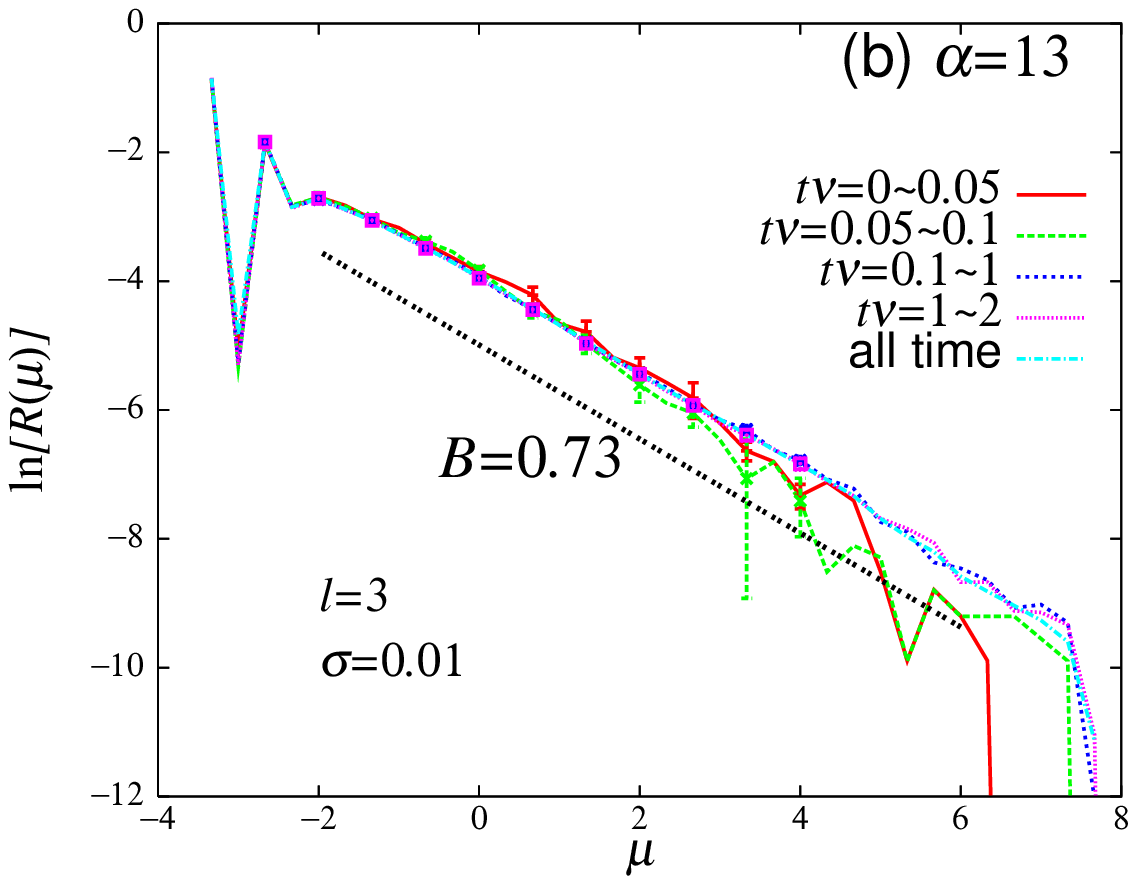}
\includegraphics[scale=0.65]{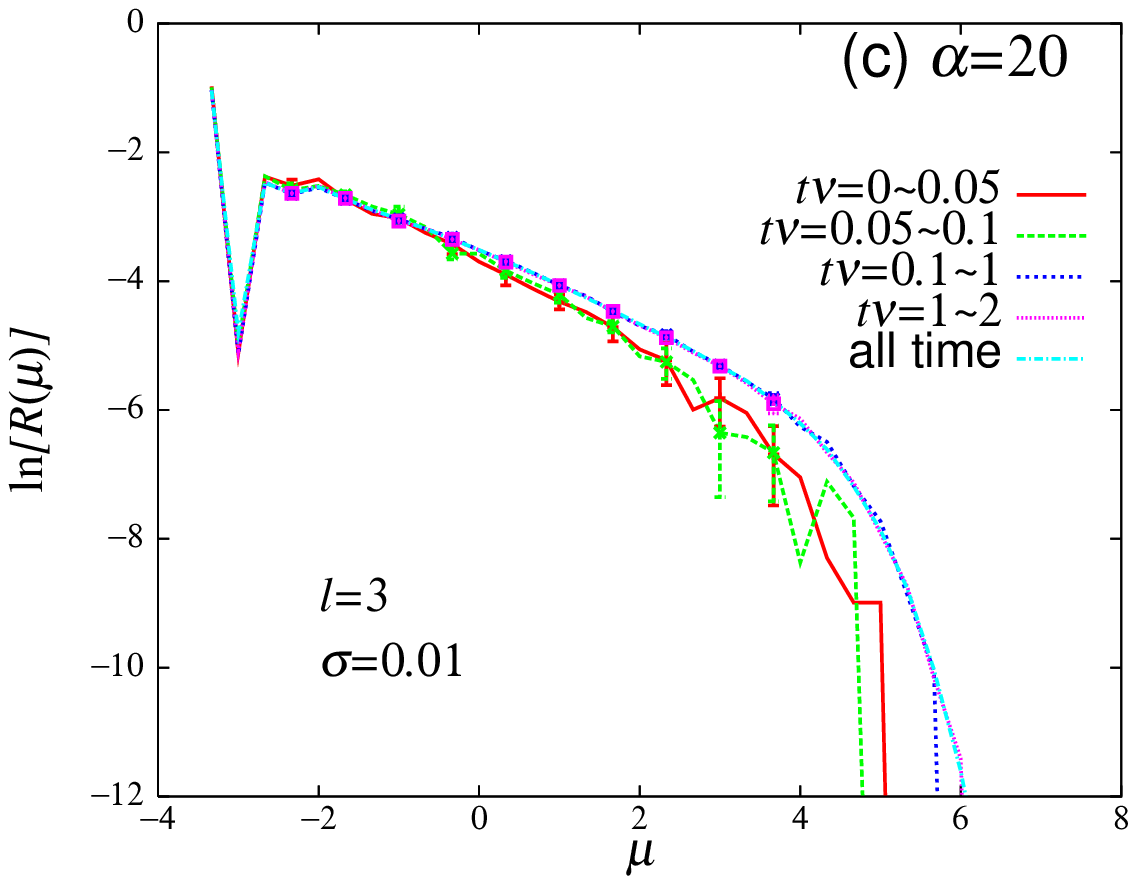}
\end{center}
\caption{
Local magnitude distribution before the mainshock with $\mu >\mu _c=5$, for several time periods before the mainshock, for the cases of
$\alpha=1$ (a), $\alpha=13$ (b) and $\alpha=20$ (c), each corresponding to the ``supercritical'', ``near-critical'' and ``subcritical'' regimes, respectively. Events whose epicenter lies within 5 blocks from the epicenter of the upcoming mainshock are counted.
The parameters $l$ and $\sigma$ are fixed  to be $l=3$ and $\sigma =0.01$. The system size is $160\times 80$.  In Fig.(a)/(c), an apparent $B$-value decreases/increases before the mainshock, while it stays almost unchanged in Fig.(b).
}
\end{figure}

\setfigurenum{16}
\begin{figure}[ht]
\begin{center}
\includegraphics[scale=0.65]{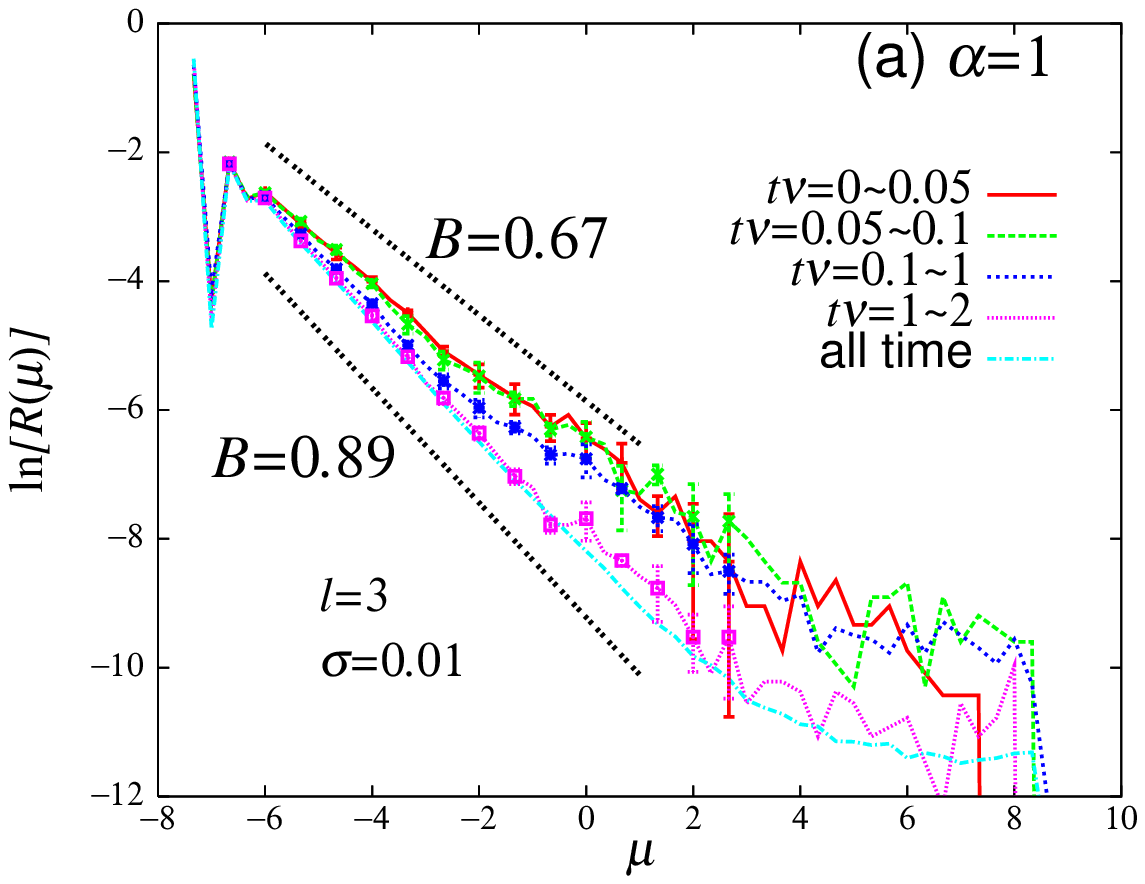}
\includegraphics[scale=0.65]{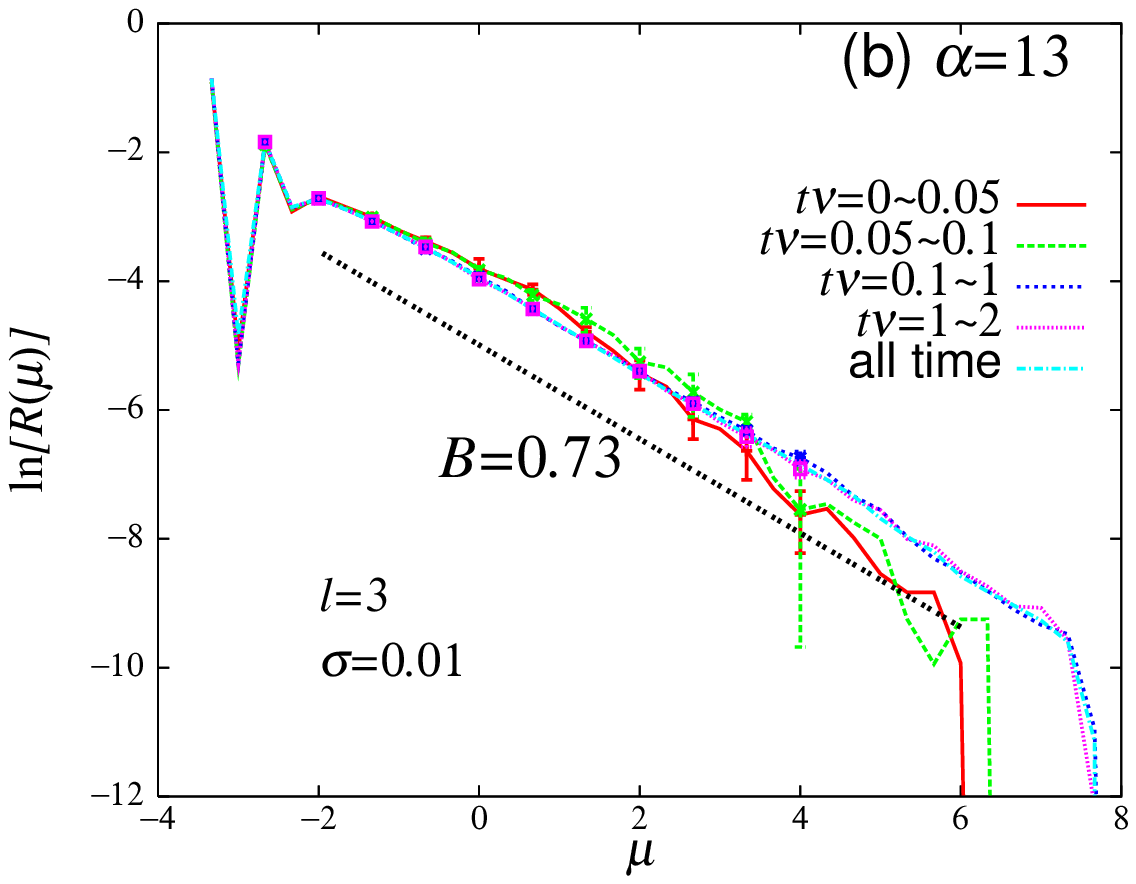}
\includegraphics[scale=0.65]{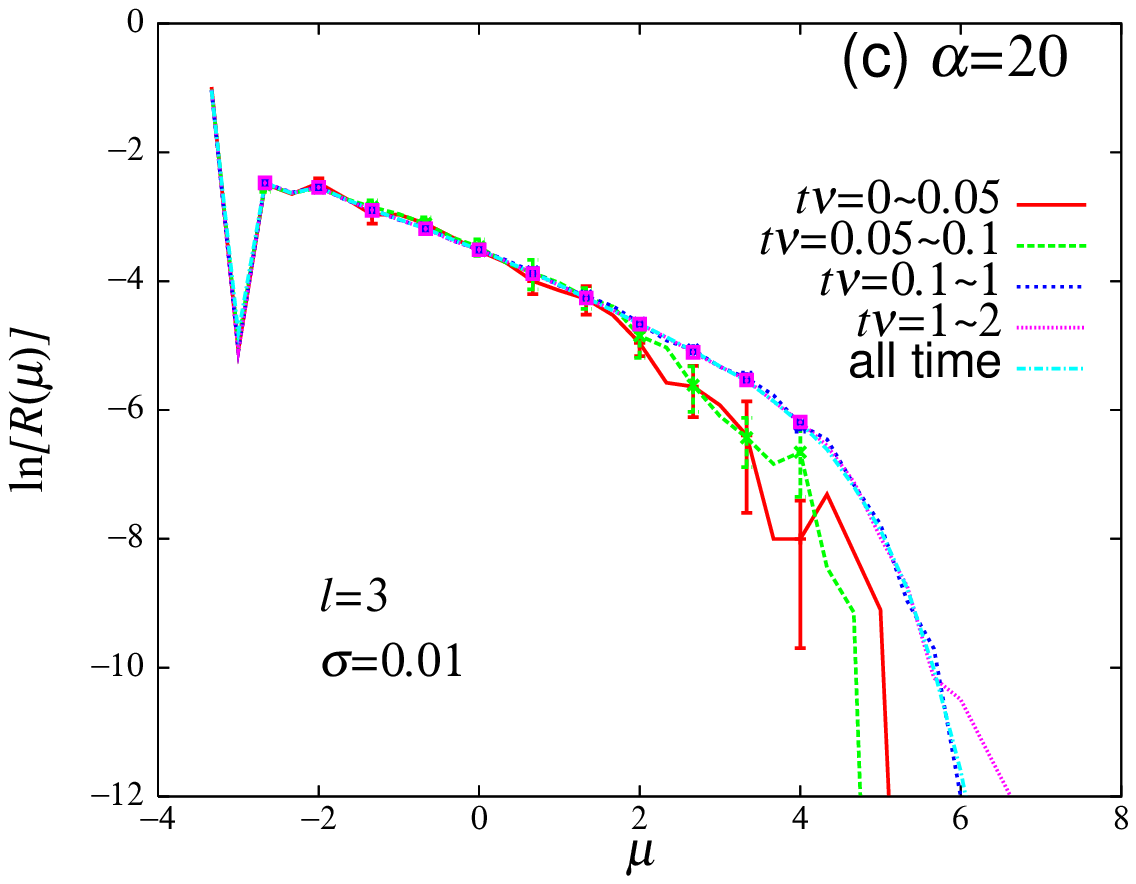}
\end{center}
\caption{
Local magnitude distribution after the mainshock with $\mu >\mu _c=5$, for several time periods after the mainshock, for the cases of
$\alpha=1$ (a), $\alpha=13$ (b) and $\alpha=20$ (c), each corresponding to the ``supercritical'', ``near-critical'' and ``subcritical'' regimes, respectively.
Events whose epicenter lies within 5 blocks from the epicenter of the preceding mainshock are counted. The parameters $l$ and $\sigma$ are fixed  to be $l=3$ and $\sigma =0.01$.  The system size is $160\times 80$.  In Figs.(a)/(c), an apparent $B$-value decreases/increases after the mainshock, while it stays almost unchanged in Fig.(b).
}
\end{figure}

\newpage
\end{article}

\end{document}